\documentclass[aps,prl,longbibliography,twocolumn,superscriptaddress]{revtex4-2}
\usepackage{amsmath,amssymb,bm,graphicx,color,gensymb,bbold,hyperref,keyval,url,latexsym}
\usepackage{CJK}
\usepackage[dvipsnames,svgnames,table]{xcolor}
\usepackage{enumitem}
\newcommand{\bcao}{BaCo$_2$(AsO$_4$)$_2$}
\usepackage{hyperref}
\hypersetup{
    colorlinks=true,    % false: boxed links; true: colored links
    linkcolor=NavyBlue, % color of internal links
    citecolor=NavyBlue,   % color of links to bibliography
    filecolor=NavyBlue, % color of file links
    urlcolor=NavyBlue   % color of external links
}

\begin{document}

% ==============================================================================
\title{BaCo$_2$(AsO$_4$)$_2$: Strong Kitaev, After All}
% ==============================================================================

% ==============================================================================
\author{Pavel A. Maksimov}
%\email{maksimov@theor.jinr.ru}
\affiliation{Bogolyubov Laboratory of Theoretical Physics, Joint Institute for Nuclear Research, Dubna, Moscow region 141980, Russia}
\affiliation{M. N. Mikheev Institute of Metal Physics of Ural Branch of Russian Academy of Sciences, S. Kovalevskaya St. 18, 620990 Yekaterinburg, Russia}
\begin{CJK*}{UTF8}{}
\author{Shengtao Jiang (\CJKfamily{gbsn}蒋晟韬)}
\affiliation{Stanford Institute for Materials and Energy Sciences, SLAC National Accelerator Laboratory and Stanford University, Menlo Park, California 94025, USA}
\author{L. P. Regnault} 
\affiliation{Institut Laue Langevin, 71 avenue des Martyrs, CS 20156, 38042 Grenoble Cedex 9, France}
\affiliation{Laboratoire de Magn{\'e}tisme et Diffraction Neutronique, CEA-Grenoble, 17 rue des Martyrs, 38054 Grenoble Cedex 9, France}
\author{A. L. Chernyshev}
%\email{sasha@uci.edu}
\affiliation{Department of Physics and Astronomy, University of California, Irvine, California 92697, USA}
% ==============================================================================
\date{\today}
% ==============================================================================
\begin{abstract}
The inelastic neutron scattering results and their analysis unequivocally point to a dominant Kitaev interaction in the  honeycomb-lattice cobaltate~\bcao. Our anisotropic-exchange model  closely describes {\it all} available neutron scattering data  in the material's field-polarized phase. The density-matrix renormalization group results for our model are in close accord with the unusual double-zigzag magnetic order and  the low in-plane saturation field  of~\bcao. 
\end{abstract}
% ==============================================================================
\maketitle
\end{CJK*}
% ==============================================================================
BaCo$_2$(AsO$_4$)$_2$, hereafter BaCAO---a member of the family of layered honeycomb-lattice transition-metal  compounds~\cite{Eymond69}  originally studied as possible realizations of the two-dimensional (2D) $XY$ model~\cite{LP77,LP78,LP79,LP80,LP83a,LP86,LP83,LP84}---has  been a mystery:  see Ref.~\cite{LP90} for an earlier review. Its magnetic order remains discrepant, with the classical  modeling yielding a spiral state but failing to describe its excitation spectrum~\cite{LP83,LP84,LP90,Rastelli79}, and the more recent  neutron-polarization analysis proposing an unusual double-zigzag state instead~\cite{LP06,LP18}. Equally perplexing has been the exceptionally low in-plane critical field to a magnetically-polarized state~\cite{LP79,LP90}, suggesting that BaCAO resides on the brink of a quantum phase transition. 

This enigmatic track record has recently propelled BaCAO into being considered a candidate for proximity to an exotic spin-liquid state~\cite{Cava_2020_BaCo}, as the search for a realization of the Kitaev model~\cite{KITAEV2006} has shifted  from  compounds of $5d$ Ir and $4d$ Ru elements~\cite{Jackeli,Winter_review,us_PRR,aRu_saga} to  $3d$ cobaltates~\cite{Giniyat_Co_2018,Motome_Co_2018,Khaliullin_3d_2020,KimKim21,Winter_Co_2022}. For $d^7$ Co$^{2+}$ ions in an edge-sharing octahedral environment, spin-orbit coupling yields $j_\text{eff}\!=\!1/2$ moments and can lead to strong Kitaev exchange via additional active orbitals and hopping paths~\cite{Khaliullin_3d_2020,KimKim21,Winter_Co_2022}. However, most  Co$^{2+}$ compounds exhibit  ferromagnetic (FM) or zigzag (ZZ) order , with  BaCAO being a stark exception~\cite{KimKim21,Winter_Co_2022,Cava_Co_2007,Songvilay_Co_2020,Ma_Co_2020,Park_Co_2020,%
NCTO_Li_2021,Li_Co_2022,Zvereva_Co_2023,Jiao_2024,fava2020,Gallegos_2024,SCGO,others}.

Recent studies on many aspects of BaCAO, such as spin excitations, phase diagram, thermal transport, and magnetoelastic properties~\cite{Armitage_THz_2021,Wang_THz_2021,BCAO_Raman_2024,Broholm_BCAO,%
Cava_2020_BaCo,Budko_2022,Zapf_2024,Tsirlin_BCAO_2024,ShiyanLi_BaCAO, Ong_BCAO_kappa_2024}, have all confirmed its status as an unusual material, but have not provided clear resolution on its model~\cite{Winter_Co_2022}. This is despite the smallness of the in-plane critical fields that make the fully polarized FM state easily accessible~\cite{LP90,Cava_2020_BaCo}, allowing one to extract BaCAO's model parameters from the inelastic neutron scattering (INS) spin-flip spectra~\cite{RaduCsCuCl,Mourigal1D,MM1,Ross11,RaduYbTiO,JeffYbTiO,CoNb,CsCeSe}.

Earlier such attempts that relied on the easy-plane ${\sf J}_1$--${\sf J}_2$--${\sf J}_3$ $XXZ$ model led to a conundrum: the extracted parameters that fit the high-field INS data did not correspond to the expected spiral ground state, nor was the model able to explain the observed fully gapped zero-field spectrum even when assuming such a state~\cite{LP84,LP90}.  

A recent study provided more INS data~\cite{Broholm_BCAO}, but rejected substantial  Kitaev terms in  BaCAO in favor of largely the same easy-plane ${\sf J}_1$--${\sf J}_3$ $XXZ$ model, imposing the zero-field order as a classical spiral state~\cite{Rastelli79}. This model yielded only a rough fit to a subset of the high-field INS data and could not reconcile   puzzling low critical fields. The growing enigma became even more bewildering as recent theoretical works suggested a complete absence of the spiral phase in the phase diagram of the quantum $S\!=\!1/2$ ${\sf J}_1$--${\sf J}_3$ $XXZ$ model~\cite{Arun_honeycomb23,shengtao_j1j3}. 

However, the easy-plane ${\sf J}_1$--${\sf J}_3$ description of BaCAO has recently received nearly unanimous~\footnote{One exception is Ref.~\cite{proximity_2023} by one of the present authors, which uses the same easy-plane $XXZ$ description of the nearest-neighbor interactions, but introduces strong bond-dependent terms in the third-neighbor exchange.} support from several first-principles  approaches~\cite{Paramekanti_Co_2021,BCAO_minimal,Kee_Co_2023,Zapf_2024}. Thus, despite its inadequate account of BaCAO's phenomenology,  the view of BaCAO as a more ordinary easy-plane ferro-antiferromagnet seemed to be  prevailing. Until now. 

In this Letter, we present INS data together with analytical and numerical results,  which unequivocally point  to a {\it dominant} Kitaev term in BaCAO's model. This model closely describes the {\it entire corpus} of available INS data for BaCAO's spectrum above the critical field, demonstrates its proximity to a phase boundary, identifies BaCAO's ground state as the non-classical double-zigzag (dZZ) state with an out-of-plane tilt of spins in accord with neutron polarimetry~\cite{LP06,LP18}, and confirms BaCAO's low in-plane critical fields.
 
% ==============================================================================
\begin{figure}[t]
\includegraphics[width=\columnwidth]{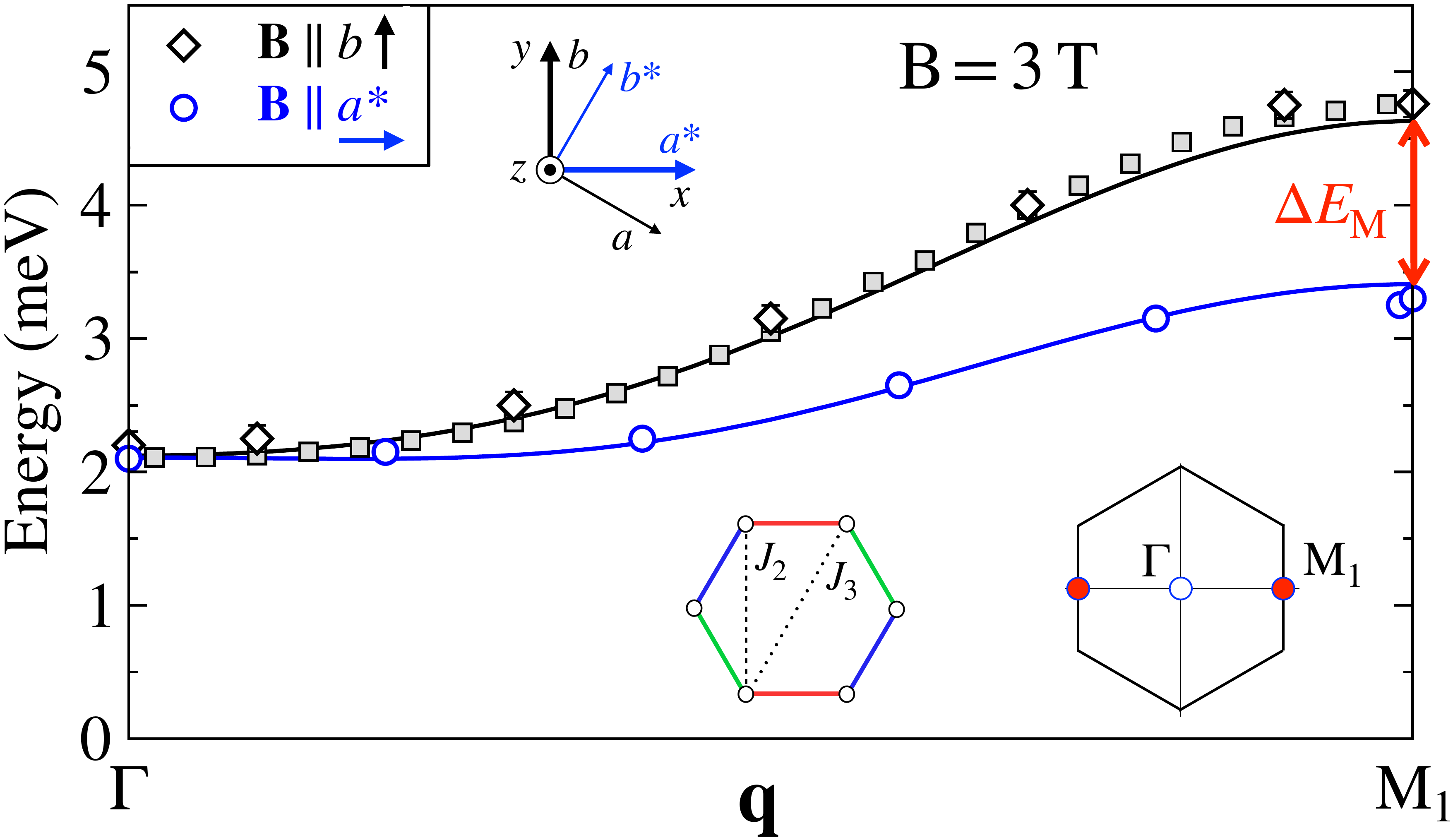}
\vskip -0.2cm
\caption{Magnon energy vs ${\bf q}$  for ${\bf B}\!\parallel\! b$ and ${\bf B}\!\parallel\! a^*$; $B\!=\!3$~T.  Diamonds~\cite{LP84}, squares~\cite{Broholm_BCAO}, and circles (this work) are the INS data, and solid lines are theory fits. $\Delta E_{\rm M}$ is the energy splitting at  M$_1$. Insets: Crystallographic axes and principal  directions, 1st-, 2nd-, and 3rd-neighbor bonds on the honeycomb lattice with  bond-dependent exchanges represented by colors, and the  Brillouin zone with $\Gamma$ and M$_1$ points indicated.}
\label{fig_lattice}
\vskip -0.5cm
\end{figure}
% ==============================================================================

{\it Key argument.}---%
While truly indispensable for inferring magnetic couplings, the spectra in the polarized phase  of Kitaev magnets can still be  challenging to analyze. Specifically, the spin-flip energies along the high-symmetry momentum directions are often independent of the very bond-dependent terms that these spectra are expected to identify~\cite{CsCeSe,GangChen16,Avers21}. For data sets dominated by such directions, it can be tempting to conclude that such terms are simply absent from the model. 

This is the case with the prior INS high-field results for BaCAO~\cite{LP84,LP90,Broholm_BCAO}. All  the existing data are for fields along one principal in-plane direction, ${\bf B}\!\parallel\! b$, perpendicular to the  bond of the honeycomb lattice. Much of the results are for momentum along the high-symmetry  $\Gamma$M$_1$ line in the Brillouin zone (BZ), in the plane normal to the field; see insets in Fig.~\ref{fig_lattice}. This is precisely the direction for which the spectrum is independent of one of the key bond-dependent Kitaev-like terms. 

Although the BaCAO spectra for  other momentum directions for that field orientation were also probed, their analyses were either unaware of Kitaev terms~\cite{LP90} or did not consider these data separately from those in the $\Gamma$M$_1$ direction to infer such terms~\cite{Broholm_BCAO}. In these studies,  easy-plane $XXZ$  models were used 
\vskip -0.15cm
\noindent
\begin{equation}
\label{eq_HXXZ}
\mathcal{H}\!=\!\sum_{n,\langle ij \rangle_n} \!{\sf J}_n \Big(S^{x}_i S^{x}_j\!+\!S^{y}_i S^{y}_j\!+\!\Delta_n S^{z}_i S^{z}_j\Big)\!-\!\sum_i \mathbf{H}\!\cdot\!\mathbf{\hat{g}}\! \cdot\!\mathbf{S}_i,
\end{equation}
\vskip -0.2cm
\noindent
where $\langle ij \rangle_n$ for $n\!=\!1,2,3$ are the $n$th-neighbor bonds, ${\sf J}_n$ are the exchanges (${\sf J}_{1,2}\!<\!0$, ${\sf J}_3\!>\!0$), and $\Delta_n$ are the $XXZ$ anisotropy parameters  for such bonds. In BaCAO, ${\sf J}_2$ is weaker than ${\sf J}_3$, as is typical in the honeycomb-lattice magnets~\cite{LP90,Winter_review}, and was neglected in Ref.~\cite{Broholm_BCAO}. All $\Delta_n$ are easy-plane-like, ${\bf H}\!=\!\mu_B {\bf B}$, and the g-tensor is diagonal, $\mathbf{\hat{g}}\!=\!\{g_x,g_y,g_z\}$, with $g_x\!\approx\!g_y\!\approx \!5.0$ and $g_z\!\approx\!2.5$~\cite{LP90}, where we use crystallographic frame in which $x\!=\!a^*$ and $y\!=\!b$; see inset of Fig.~\ref{fig_lattice}.

Because of the $U(1)$ symmetry of the $XXZ$ model (\ref{eq_HXXZ}), one does not expect any dependence of the spectra in the polarized phase on the in-plane field orientation.  

Our Figure~\ref{fig_lattice} presents  key and unambiguous evidence that the description of BaCAO with  model (\ref{eq_HXXZ}) is incomplete and that very significant bond-dependent terms must be present in it. Diamonds and squares in Fig.~\ref{fig_lattice} are the prior data from Refs.~\cite{LP84} and \cite{Broholm_BCAO}, respectively, for the spin-flip energy along the $\Gamma$M$_1$ line for the ${\bf B}\!\parallel\! b$ orientation and $B\!=\!3$~T, well above the critical field of $B_c\!\approx\!0.55$~T~\cite{LP90}. Crucially, Fig.~\ref{fig_lattice}  presents  previously unpublished results (circles), obtained at IN22 spectrometer at the Institut Laue Langevin, Grenoble, for the same momentum direction and field value, but for  a different principal in-plane field orientation, ${\bf B}\!\parallel\! a^*$; see End~Matter (EM) for the experimental details. Contrary to the expectations from the $XXZ$ model, the two bands for different field orientations show a large splitting. 

The demonstrated energy splitting $\Delta E_{\rm M}$ at the M$_1$ point in Fig.~\ref{fig_lattice} is about half of the magnon's bandwidth, suggesting that the Kitaev-like exchanges should be on par with the $XXZ$ terms of  model (\ref{eq_HXXZ}). The solid lines use the model discussed below, which will make the magnitude of these terms precise and explicit. 

% ==============================================================================
\begin{figure}[t]
\includegraphics[width=\columnwidth]{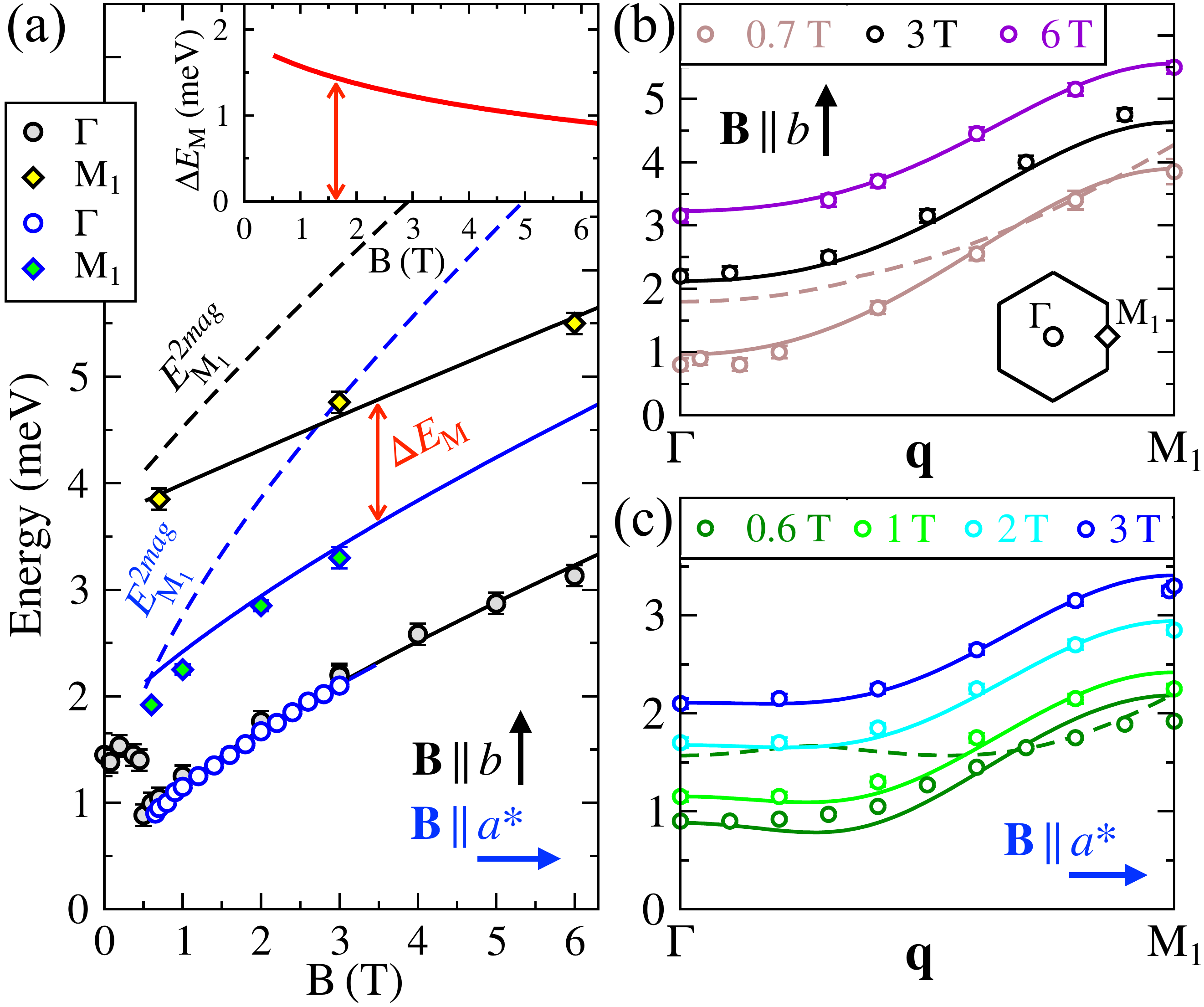}
\vskip -0.22cm
\caption{(a) The field dependence of $E_\Gamma$ (circles) and $E_{\rm M_1}$ (diamonds)  for ${\bf B}\!\parallel\! b$ and ${\bf B}\!\parallel\! a^*$, shown as black and blue symbols (data) and lines (theory), respectively. $\Delta E_{\rm M}$ is indicated and shown in the inset. (b) and (c), Same as in Fig.~\ref{fig_lattice}: data and their fits for magnon energies vs ${\bf q}$  with ${\bf B}\!\parallel\! b$,  $B\!=\!0.7$~T, 3.0~T, and 6.0~T from Refs.~\cite{LP84,LP90} in (b), and with ${\bf B}\!\parallel\! a^*$, $B\!=\!0.6$~T, 1.0~T, 2.0~T, and 3.0~T (this work) in (c). Dashed lines represent the lowest edges of the two-magnon continua for 0.7~T in (b),  0.6~T in (c), and for the  ${\rm M_1}$ point in (a).}
\label{fig_INS}
\vskip -0.5cm
\end{figure}
% ==============================================================================

The large band splitting for the two field orientations is further exposed in Figure~\ref{fig_INS}(a), which shows the field dependence of the energies at the $\Gamma$ and M$_1$ points.  While $E_\Gamma$ data (circles) are nearly degenerate for ${\bf B}\!\parallel\! b$ and ${\bf B}\!\parallel\! a^*$ (black and blue color, respectively), indicating  $g_{b}\!\approx\!g_{a^*}$,  $E_{\rm M_1}$ results (diamonds) exhibit a large gap $\Delta E_{\rm M}$ between the two orientations, which is maximal at lower fields. Solid lines show  theoretical fits of the  $E_{\Gamma(\rm M_1)}$ data, and the inset shows the fit of the gap $\Delta E_{\rm M}$ vs field.

 % ==============================================================================
\begin{figure*}
\centering
\includegraphics[width=\linewidth]{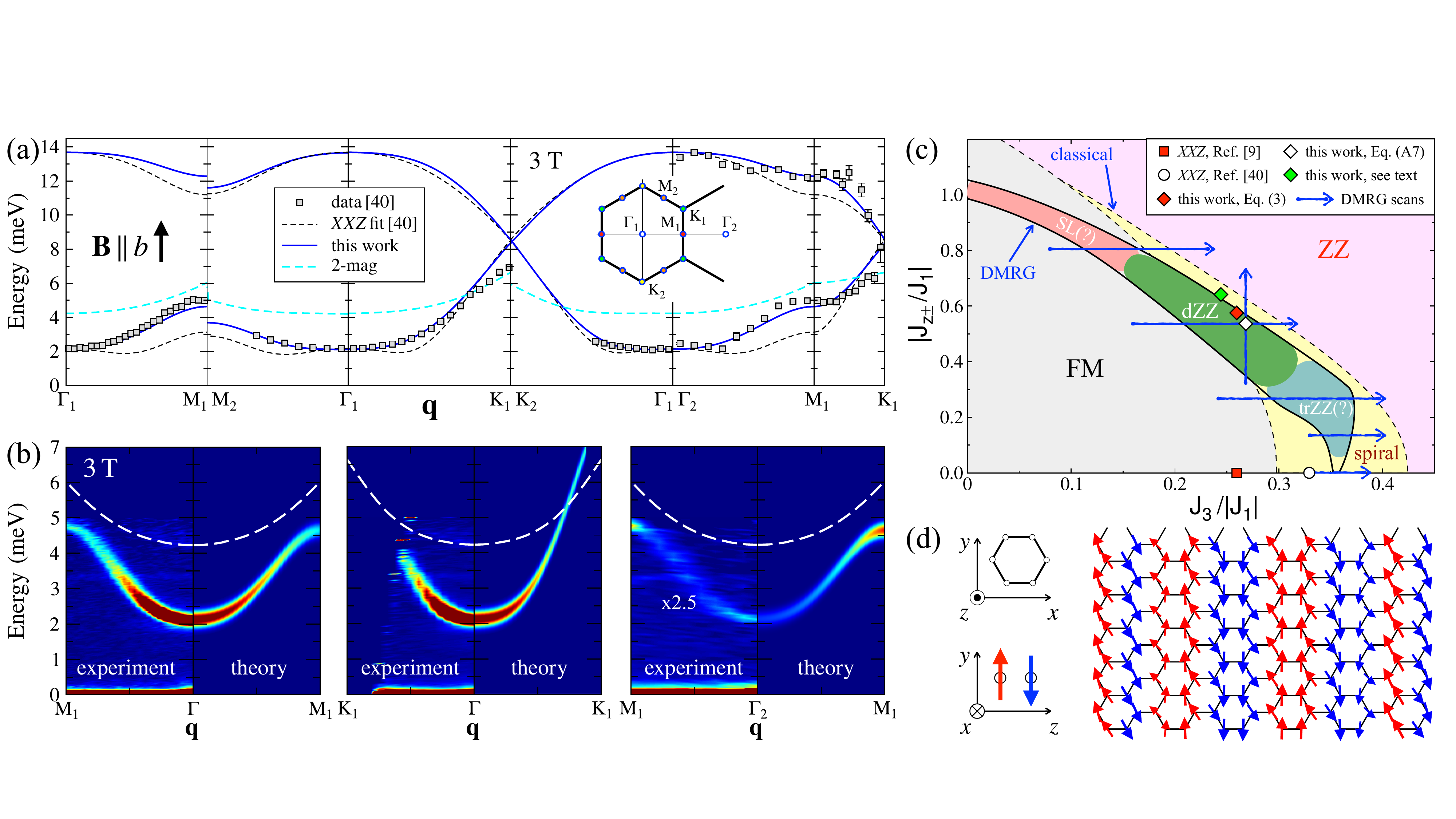}
\vskip -0.2cm
\caption{(a) Data from Ref.~\cite{Broholm_BCAO} for ${\bf B}\!\parallel\! b$, $B\!=\!3$~T, lines are as described in the legend. (b) Intensity maps of ${\cal S}({\bf q},\omega)$ from Ref.~\cite{Broholm_BCAO} and our results. Dashed lines mark the bottom of the two-magnon continuum. (c) The ${\sf J}_{z\pm}$--${\sf J}_3$ phase diagram of the model (\ref{eq_HXXZ})$+$(\ref{eq_HJpm}) with the LT and DMRG  boundaries and phases; $\Delta_n$, ${\sf J}_2/{\sf J}_1$, and ${\sf J}_{\pm\pm}/{\sf J}_1$ are from (\ref{eq_set_Jpp}), symbols show various parameter sets (see the text). (d) Results from DMRG $12\!\times\!12$ cluster calculations for the set (\ref{eq_set_Jpp}) with spins  in the $yz$ plane.}
\label{fig_spectrum}
\vskip -0.5cm
\end{figure*}
% ==============================================================================

In Fig.~\ref{fig_INS}(b), we reproduce the INS data from Refs.~\cite{LP84,LP90} for the magnon energies along the $\Gamma$M$_1$ line for ${\bf B}\!\parallel\! b$ with $B\!=\!0.7$~T, 3.0~T, and 6.0~T.  Fig.~\ref{fig_INS}(c) presents previously unpublished results for ${\bf B}\!\parallel\! a^*$ with $B\!=\!0.6$~T, 1.0~T, 2.0~T, and 3.0~T, showing the field evolution of the band that remains much flatter than the one in  Fig.~\ref{fig_INS}(b). The solid lines represent the theoretical fits.

 We also highlight the two-magnon continua, with their lowest edges shown by  dashed lines in  Figs.~\ref{fig_INS}(b) and \ref{fig_INS}(c) for the fields of 0.7~T and 0.6~T, respectively. These lines demonstrate the overlap of the continua with the single-magnon branches that may be responsible for the deviations of the observed energies from the theoretical fits. Dashed lines in Fig.~\ref{fig_INS}(a) serve the same purpose. 

{\it Model parameters.}---%
The symmetry-allowed bond-dependent terms in the 1st-neighbor exchange matrix of BaCAO~\cite{Winter_Co_2022}, missing from the $XXZ$ model (\ref{eq_HXXZ}), are
\vskip -0.1cm
\noindent
\begin{align}
&{\cal H}_{bd}\!=\!\sum_{\langle ij\rangle_1} \!\Big\{{\sf J}_{z\pm}\Big(\big( S^{x}_i S^{z}_j \!+\!S^{z}_i S^{x}_j \big) \tilde{s}_\alpha \!-\!\big( S^{y}_i S^{z}_j\!+\!S^{z}_i S^{y}_j\big)\tilde{c}_\alpha \Big)\nonumber\\[-5pt]
\label{eq_HJpm}
&\ \ \ +\! 2 {\sf J}_{\pm \pm} 
\Big( \big( S^{x}_i S^{x}_j \!- \!S^{y}_i S^{y}_j \big) \tilde{c}_\alpha 
\!-\!\big( S^{x}_i S^{y}_j\!+\!S^{y}_i S^{x}_j\big)\tilde{s}_\alpha \Big)\Big\},
\end{align}
\vskip -0.15cm
\noindent
where $\tilde{c}(\tilde{s})_\alpha\!=\!\cos(\sin)\tilde{\varphi}_\alpha$ and $\tilde{\varphi}_\alpha\!=\!\{0,2\pi/3,-2\pi/3\}$ are the bond angles with the $x$ axis in Fig.~\ref{fig_lattice}.  The Kitaev exchange is given by $K\!=\!\sqrt{2}{\sf J}_{z\pm}\!-2{\sf J}_{\pm\pm}$. For the full translation of the $XXZ$ and (\ref{eq_HJpm}) model to the Kitaev notations, $\{ {\sf J}_1,\Delta_1{\sf J}_1,{\sf J}_{\pm \pm},{\sf J}_{z\pm}\}\!\Leftrightarrow\!\{J,K,\Gamma,\Gamma'\}$; see EM~\cite{aRu_saga,Winter_review}. 

It is the ${\sf J}_{z\pm}$ term in (\ref{eq_HJpm}) to which the $\Gamma$M$_1$ INS data for the high ${\bf B}\!\parallel\! b$ fields are  insensitive, making it difficult to detect~\cite{CsCeSe}. It is also the only term in the model (\ref{eq_HXXZ})$+$(\ref{eq_HJpm}) that can induce a tilt of spins out of the crystallographic plane, as it is the only one that couples the in-plane and  out-of-plane spin components~\cite{us_PRR,aRu_saga}. 

BaCAO's magnetization data suggest that its field-induced state, from just above the in-plane critical field $H_c\!\approx\!0.55$~T,  is  nearly fully polarized and free from quantum fluctuations ~\cite{Broholm_BCAO,Wang_THz_2021,ShiyanLi_BaCAO,Cava_2020_BaCo}. Therefore, one can extract exchange parameters from the INS spin-flip spectra using  spin-wave theory (SWT) for such a state. 

Our model contains seven terms:  symmetry-allowed nearest-neighbor exchanges, $\{ {\sf J}_1,\Delta_1,{\sf J}_{\pm \pm},{\sf J}_{z\pm}\}$,   third-neighbor $XXZ$ couplings,  $\{ {\sf J}_3,\Delta_3\}$, and a small  ${\sf J}_2$~\cite{LP90}, which is kept  $XY$-like, $\Delta_2\!=\!0$, to avoid overfitting. We also used slightly different in-plane g-factors, $g_{a*}\!=\!4.8$ and $g_b\!=\!4.85$, to fit the field-dependencies of the energies at the $\Gamma$ and M points  in Fig.~\ref{fig_INS}.

We use the INS data for the reference fields of 3~T and 6~T to fix different combinations of the exchanges by the energies at  high-symmetry ${\bf q}$-points, such as $\Gamma$ and M-points in Figs.~\ref{fig_lattice} and \ref{fig_INS}. Even at 3~T, the two-magnon continuum can cross the single-magnon branch and affect the data, as can be seen in Fig.~\ref{fig_spectrum}(a) for the proximity of the K-point. This makes the use of straightforward quality fit criteria more problematic,  because  the SWT K-point energy needs to be higher than the data. 

Delegating further technical details of the parameters' extraction to  Supplemental Material (SM)~\cite{SM}, the proposed BaCAO parameters (in meV, except for $\Delta_n$) are
\vskip -0.15cm
\noindent
\begin{align}
\{{\sf J}_1, \Delta_1, {\sf J}_{\pm\pm}, {\sf J}_{z\pm}\}=\{-6.54, 0.36,0.15, -3.76\}
\label{eq_set_Jpp}
\end{align}
\vskip -0.2cm
\noindent
and $\{{\sf J}_2,{\sf J}_3,\Delta_3\}\!=\!\{-0.21, 1.70,0.03\}$.

This is the set of parameters that provides a close fit to {\it all} available INS data for BaCAO in the polarized phase, as shown in Figs.~\ref{fig_lattice} and \ref{fig_INS}, and Figs.~\ref{fig_spectrum}(a) and \ref{fig_spectrum}(b) for ${\bf B}\!\parallel\! b$ at $B\!=\!3$~T with the data from Ref.~\cite{Broholm_BCAO}. In Fig.~\ref{fig_spectrum}(a), we show the data in multiple ${\bf q}$-panels together with the $XXZ$ fit from Ref.~\cite{Broholm_BCAO} (black dashed lines) and our results for the single-magnon branches and the bottom of the two-magnon continuum (solid and light-blue dashed lines, respectively).  Fig.~\ref{fig_spectrum}(b) shows a side-by-side comparison of the intensity maps of the dynamical structure factor ${\cal S}({\bf q},\omega)$ from Ref.~\cite{Broholm_BCAO} and our theoretical results. The comparison of our results with the data for $B\!=\!0.75$~T  and ${\bf B}\!\parallel\! b$ from Ref.~\cite{Broholm_BCAO} is shown in EM.

We note that the $XXZ$ sector of our BaCAO parameters in (\ref{eq_set_Jpp}) is close to those in earlier studies, Refs.~\cite{LP84,LP90}, with $\Delta_1\!\approx\!0.4$, ${\sf J}_3/|{\sf J}_1|\!\approx\!0.26$, and ${\sf J}_2/{\sf J}_1\!\approx\!0.034$ compared to our $\Delta_1\!=\!0.36$, ${\sf J}_3/|{\sf J}_1|\!=\!0.26$, and ${\sf J}_2/{\sf J}_1\!=\!0.032$~\footnote{The absolute values of exchanges in Refs.~\cite{LP84,LP90} used an additional factor of 2 in their conventions}. Since they describe the earlier INS data well~\cite{SM}, it suggests that ${\sf J}_{\pm\pm}$ in (\ref{eq_HJpm}) should be small, as is the case in our proposed set (\ref{eq_set_Jpp}). A large ${\sf J}_{\pm\pm}$ would also yield a field-dependence of energies for ${\bf B}\!\parallel\! a^*$ that is incompatible with the new INS data~\cite{SM}. This leaves a necessarily large ${\sf J}_{z\pm}$ as the only term capable of explaining the $\Delta E_M$ gap in Figs.~\ref{fig_lattice} and \ref{fig_INS}(a), which is our key result.

Last, but not least, translating the nearest-neighbor exchanges in (\ref{eq_set_Jpp}) to the Kitaev notations yields 
\vskip -0.1cm
\noindent
\begin{align}
\{J, K, \Gamma, \Gamma'\}\approx\{-3.3, -5.6, 3.0, 0.6\}\text{ meV},
\label{eq_set}
\end{align}
\vskip -0.15cm
\noindent
with a dominant Kitaev term, approximately twice the magnitude of either $J$ or $\Gamma$. 

{\it Phase diagram.}---%
Our parameters  for BaCAO (\ref{eq_set_Jpp}) also resolve the enigma of its ground state and identify its place in a broader phase diagram of Kitaev magnets. 

In Fig.~\ref{fig_spectrum}(c), we present the classical Luttinger-Tisza (LT) ${\sf J}_{z\pm}$--${\sf J}_3$ phase diagram of the model (\ref{eq_HXXZ})$+$(\ref{eq_HJpm}) for $\Delta_1\!=\!0.36$ and ${\sf J}_2/{\sf J}_1\!=\!0.03$ as in (\ref{eq_set_Jpp}), and we  set  $\Delta_3\!=\!{\sf J}_{\pm\pm}\!=\!0$ for simplicity~\cite{aRu_saga,lt_original}. The sequence of the  FM, ZZ, and  helical phases shown in Fig.~\ref{fig_spectrum}(c) is common in many classical honeycomb-lattice models~\cite{Rastelli79,Arun_honeycomb23,aRu_saga}. 

The red square and white circle represent projections of the $XXZ$ models of BaCAO from Refs.~\cite{LP84,LP90} and Ref.~\cite{Broholm_BCAO}, respectively, with the first in the FM phase and the second matching the observed ordering vector ${\bf k}\!\approx\!(0.27,0)$~r.l.u.~as a classical spiral~\cite{Broholm_BCAO}.  Our proposed set (\ref{eq_set_Jpp}) is represented by the red diamond. Classically, it also corresponds to  a spiral with ${\bf k}\!\approx\!(0.16,0)$~r.l.u.~and small in-plane critical fields $\alt\!0.15$~T. The  green diamond  represents a set with similar classical characteristics and provides slightly better fits to the  ${\bf B}\!\parallel\! a^*$ INS spectra in Fig.~\ref{fig_INS}~\cite{SM}. The empty diamond represents a set that is also similar classically,  showing comparable, but slightly worse overall quality of fit to the INS data; see EM.  

Recent studies of the quantum $S\!=\!1/2$ edition of the $XXZ$ ${\sf J}_1$--${\sf J}_3$ model using the density-matrix renormalization group (DMRG) method~\cite{Arun_honeycomb23,shengtao_j1j3} found no trace of the spiral phase in its phase diagram, uncovering instead only narrow ranges of non-classical states, such as dZZ. Here, we extend this study to the ${\sf J}_1$--${\sf J}_{z\pm}$--${\sf J}_3$ model, with the remaining parameters ($\Delta_n$, ${\sf J}_{\pm\pm}$, and ${\sf J}_2$) fixed according to the proposed BaCAO set in (\ref{eq_set_Jpp}). 

In Fig.~\ref{fig_spectrum}(c),  we show the boundaries of the intermediate phases, which are inferred by  extrapolation from several DMRG ``scans,'' shown by horizontal and vertical arrows. Many individual points  were also checked by DMRG (``non-scans'') to verify the phases. DMRG calculations~\cite{itensor}, which used the same approach for a closely related model, have been performed recently~\cite{aRu_saga} with a similar goal of exploring intermediate phases between FM and ZZ; see EM and Ref.~\cite{aRu_saga} for technical details.  

Similar to  prior works~\cite{Arun_honeycomb23,shengtao_j1j3}, the spiral phase is eliminated along the $XXZ$ line. However, at finite ${\sf J}_{z\pm}$, a sequence of non-classical quasi-collinear phases emerges. At  lower ${\sf J}_{z\pm}$, the triple-zigzag state, not unlike the one briefly discussed in Ref.~\cite{shengtao_j1j3},  occurs, while for the region relevant to BaCAO, a dZZ state firmly establishes itself. For  larger ${\sf J}_{z\pm}$, we observed a state that resembles a spin liquid (SL), with only short-ranged correlations. However, further analysis of this state is needed. The set highlighted by a green diamond is  in the ZZ phase. 

DMRG simulation of the proposed parameter set (\ref{eq_set_Jpp}) on a $12\times 12$ cylinder are shown in  Fig.~\ref{fig_spectrum}(d), demonstrating  the dZZ state with spins shown in the $yz$ plane to emphasize their out-of-plane tilt. While the non-classical dZZ state for the $XXZ$ ${\sf J}_1$--${\sf J}_3$ models has recently received theoretical endorsements~\cite{Zapf_2024,proximity_2023,BCAO_minimal,shengtao_j1j3}, the tilt is unusual and is clearly due to the ${\sf J}_{z\pm}$ coupling. According to our DMRG results, every {\it second} spin in the dZZ structure is tilted out of plane by about  $35\degree$.  Using the g-factors $g_z/g_{ab}\!\approx\!0.5$, this yields $7.8\degree$ average tilt of the magnetic moment, compared with $\sim\!6\degree$ suggested by neutron polarimetry~\cite{LP06,LP18}.
These and other quantitative discrepancies can be attributed either to finite-size effects in DMRG or to small terms in the description of BaCAO that are not accounted for by our model~\cite{SM}.

As one can see in Fig.~\ref{fig_spectrum}(c), the proposed  parameter set (\ref{eq_set_Jpp}) appears close to the dZZ-ZZ phase boundary within the  dZZ phase, likely related to the anomalously low in-plane critical fields in BaCAO, $g_{ab} \mu_B B_c/|{\sf J}_1| \!\simeq\! 0.02$~\cite{LP77}.  Using the DMRG scans vs field~\cite{SM}, we have found  critical fields for our model of BaCAO slightly higher than the observed 0.55~T values:  $B_c^{(b)}\!\approx\!B_c^{(a^*)}\!\approx\!0.7$~T,  but nearly equal, in agreement with Refs.~\cite{Zapf_2024,Tsirlin_BCAO_2024}. The out-of-plane critical field obtained by the DMRG scan is  about 47~T~\cite{SM}, consistent with experiments~\cite{Zapf_2024}. Lastly, the field-induced ``up-up-down'' columnar state---present in the BaCAO's phase diagram below the polarized state---has also been captured by our model (see SM~\cite{SM}).

{\it Summary.}---%
Our study brings the multifaceted conundrum of BaCAO close to a complete resolution. The INS data presented in this work provide the most direct evidence of a dominant Kitaev term in its model, establishing BaCAO as a Kitaev-champion among cobaltates and comparable to, if not superior to, other Kitaev magnets, all while reconciling its puzzling ground state and low critical fields with its model's phenomenology.

Our nearly ideal fit of {\it all} available INS spectra in the  polarized state also includes nuances such as
analyses of the crossings with the two-magnon continuum. The ground state of our model for BaCAO corresponds to the non-classical double-zigzag state with a finite tilt of spins out of the crystallographic plane, all in accord with the neutron polarimetry data. The proposed parameter set for BaCAO is shown to correspond to near proximity to a phase boundary, with  small and nearly isotropic in-plane and large out-of-plane critical fields, all consistent with the material's phenomenology.

For the broader field of Kitaev magnets, the striking variety of phases present between the ferromagnetic and zigzag states in their models' phase diagrams calls for further in-depth studies. Inspired by the BaCAO example, explorations of materials that potentially host bond-dependent interactions should be pursued to uncover novel phases and phenomena.

%%%%%%%%%%%%%%%%%%%%%%%%%%%%%%%%%%%%%%%

% ============================================================================
\begin{acknowledgments}
{\it Acknowledgments.}---%
% ============================================================================
We are greatly indebted to T.~Halloran and C.~Broholm for sharing their published inelastic neutron scattering data. We also thank S.~Streltsov and S.~Winter for their useful discussions, and Chaebin Kim for raising an important question. 

This work was primarily supported by the U.S. Department of Energy, Office of Science, Basic Energy Sciences under Award No. DE-SC0021221 (A.~L.~C.).
Spin-wave calculations (P.~A.~M.) were supported by the Russian Science Foundation via project 23-12-00159.  The work by  S.~J. was supported by the National Science Foundation under DMR-2110041. S.~J. is also supported by the Department of Energy (DOE), Office of Sciences, Basic Energy Sciences, Materials Sciences and Engineering Division, under Contract No. DEAC02-76SF00515. 

A.~L.~C.  would like to thank Aspen Center for Physics and the Kavli Institute for Theoretical Physics (KITP) where different stages of this work were advanced. The Aspen Center for Physics is supported by National Science Foundation Grant No. PHY-2210452 and KITP is supported by the National Science Foundation under Grant No. NSF PHY-2309135.
\end{acknowledgments}
%----------------------------------------

\bibliography{bcao_bib}

%\vspace{-0.3cm}
\newpage
% ==============================================================================
%---------------------------------------------------------------------------
\onecolumngrid
%---------------------------------------------------------------------------
\begin{center}
 \vskip 0.15cm
{\large\bf End Matter}
\end{center}
\vskip 0.15cm
%---------------------------------------------------------------------------
\twocolumngrid
%---------------------------------------------------------------------------

%%%%%%%%%%%%%%%%%%%%%%%%%%%%%%%%%%%%%%%%%%%%%%%%%%
\renewcommand{\theequation}{A\arabic{equation}}
%%%%%%%%%%%%%%%%%%%%%%%%%%%%%%%%%%%%%%%%%%%%%%%%%%
\setcounter{equation}{0}
% ==============================================================================

% ==============================================================================
\begin{figure}[t]
\centering
\includegraphics[width=0.8\linewidth]{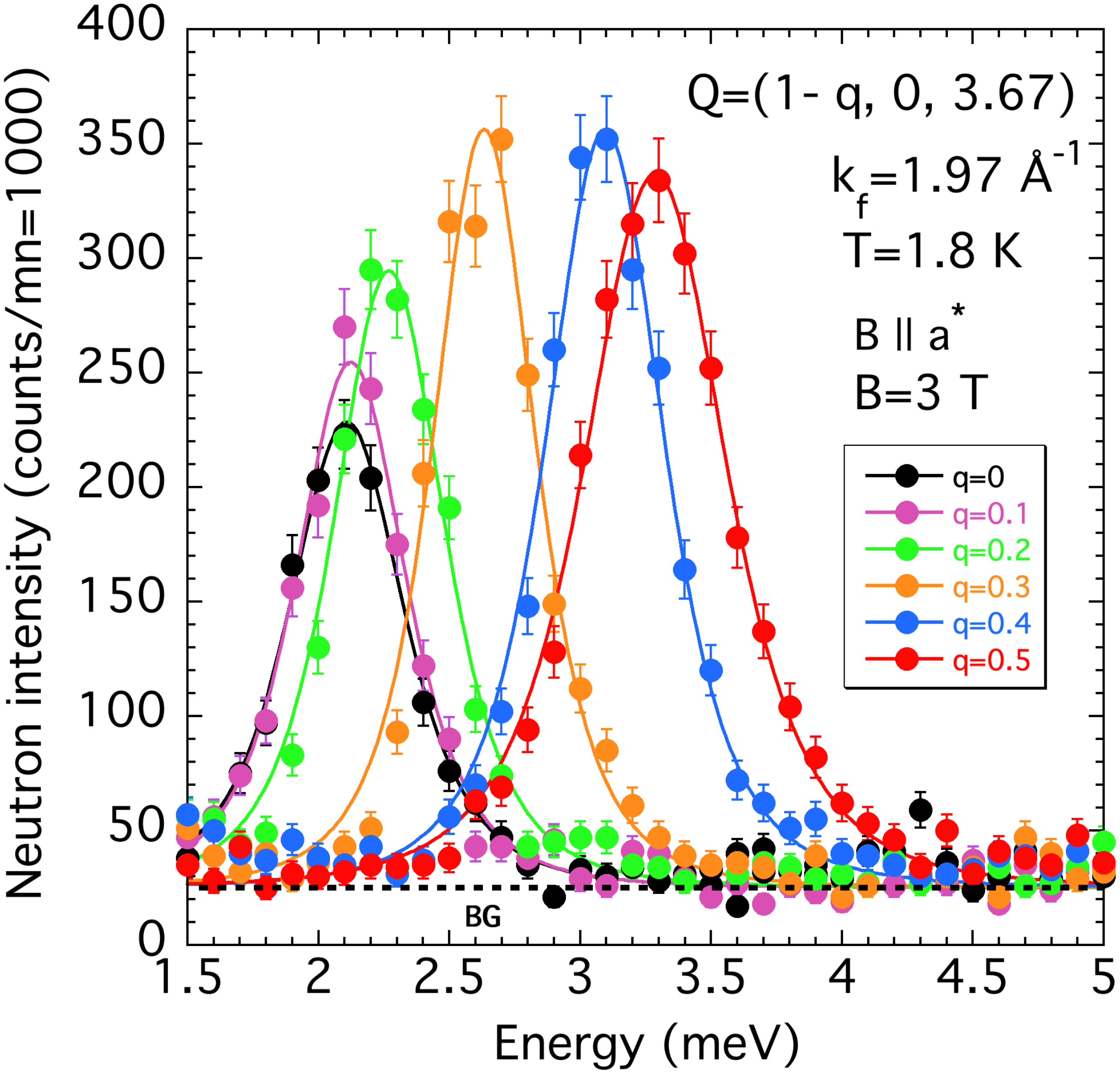}
\vskip -0.3cm
\caption{Constant-Q energy scans at the scattering vector ${\bf Q}\!=\!(0.5, 0,3.67)$, and fixed $k_f\!=\!1.97 \AA^{-1}$ for  $B\!=\!3$~T and ${\bf B}\!\parallel\!{\bf a^*}$. The peak maxima correspond to magnon energies in Figs.~\ref{fig_lattice} and \ref{fig_INS} along the $\Gamma$M$_1$ ${\bf q}$-direction. For the raw data, see Ref.~\cite{Zenodo_data}.}
\label{Fig-4}
\vskip -0.3cm
\end{figure}
% ==============================================================================

% ============================================================================
{\it INS details.}---%
% ============================================================================
The INS experiments were carried out on the three-axis spectrometer (TAS) IN22 at the Institut Laue Langevin, Grenoble,  using 3.8-T horizontal-field ($\mathbf{B}\!\parallel\! a^*$) and 5-T vertical-field cryomagnets at $T\!=\!1.8\text{~K}$. Previous measurements  had been performed in vertical field ($\mathbf{B}\!\parallel \! b$) at $T\!=\!4.2\text{~K}$ on TAS DN1 at the Siloe reactor, CEA-Grenoble~\cite{LP84}. In all cases, the sample was aligned with the $b$ axis perpendicular to the ($a^*$, $c^*$) scattering plane. Figure~\ref{Fig-4} shows typical energy scans for a magnetic field of $B\!=\!3$~T and ${\bf B}\!\parallel\!{\bf a^*}$ (see SM~\cite{SM} for more details). The peak positions correspond to the magnon energies shown in Figs.~\ref{fig_lattice} and \ref{fig_INS} along the $\Gamma$M$_1$ ${\bf q}$-direction. Here and elsewhere, the neutron transfer momentum ${\bf q}\! =\! {\bf k}_f - {\bf k}_i$ and the scattering vector ${\bf Q}\!=\!-{\bf q}$.

\begin{figure}[t]
\includegraphics[width=0.6\linewidth]{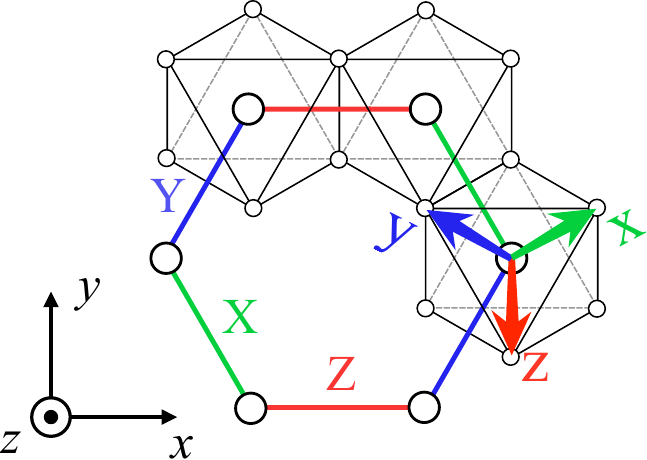}
\vskip -0.2cm
\caption{Crystallographic and cubic axes in the honeycomb-lattice structure with an octahedral environment.}
\label{FigEM:cubic_axes}
\vskip -0.3cm
\end{figure}

% ============================================================================
{\it Model details.}---%
% ============================================================================
The model \eqref{eq_HXXZ}+\eqref{eq_HJpm} is written in the crystallographic reference frame $\{x,y,z\}$, which is defined by the honeycomb plane. The nearest-neighbor extended Kitaev-Heisenberg model uses the cubic reference frame $\{\rm x,y,z\}$ shown in Fig.~\ref{FigEM:cubic_axes}, and is given by
\vskip -0.05cm
\noindent
\begin{align}
\mathcal{H}_1=&\sum_{\langle ij \rangle_\gamma} \Big[J {\bf S}_i \cdot {\bf S}_j 
+K S^\gamma_i S^\gamma_j +\Gamma \big(S^\alpha_i S^\beta_j +S^\beta_i S^\alpha_j\big)\nonumber \\[-10pt]
\label{eq_H_JKG}
&\ \ \ \ \ +\Gamma' \big(S^\gamma_i S^\alpha_j+
S^\gamma_i S^\beta_j+S^\alpha_i S^\gamma_j +S^\beta_i S^\gamma_j\big)\Big],
\end{align}
\vskip -0.05cm
\noindent
where $\{\alpha,\beta,\gamma\}=\{{\rm y,z,x}\}$ for the X bond, as shown in Fig.~\ref{FigEM:cubic_axes}, and the interactions on Y and Z bonds are obtained through cyclic permutations. The two representations of the exchange model are related via an axes rotation, with the transformation given by 
\begin{eqnarray}
\label{KJGG1_model}
&&\left(
\begin{array}{c}
J\\
K\\
\Gamma\\
\Gamma'\\
\end{array}
\right) \!=\!
\left(
\begin{array}{cccc}
 2/3 & 1/3 & 2/3 & -\sqrt{2}/3 \\
 0 & 0 & -2 & \sqrt{2} \\
 -1/3 & 1/3 & -4/3 & -\sqrt{2}/3 \\
 -1/3 & 1/3 & 2/3 & \sqrt{2}/6\\
\end{array} 
\right) \! \left(
\begin{array}{c}
{\sf J}\\
\Delta {\sf J}\\
{\sf J}_{\pm\pm}\\
{\sf J}_{z\pm}\\
\end{array}
\right)\!. \ \ \ \ \ \
\end{eqnarray}

% ==============================================================================
% ==============================================================================

There is another, fully equivalent parametrization of the nearest-neighbor exchange matrix within the same model. A global 180$\degree$ rotation of the crystallographic frame about the $z$ axis leaves the model~\eqref{eq_HXXZ}+\eqref{eq_HJpm} invariant (up to an inconsequential sign change ${\sf J}_{z\pm}\!\rightarrow\!-{\sf J}_{z\pm}$), but reshuffles the parameters $\{J,K,\Gamma,\Gamma'\}$  via a linear transformation---this relationship is known as a form of duality between one generalized Kitaev-Heisenberg model and another~\cite{dualityPRX}. The corresponding parameter set that is physically equivalent to the one used in this work (\ref{eq_set}) is given by
\vskip -0.1cm
\noindent
\begin{align}
\{J, K, \Gamma, \Gamma'\}_{\rm eqv}\approx\{-6.8, 5.0, -0.6, 2.4\}\text{ meV},
\label{eq_set_alt}
\end{align}
\vskip -0.15cm
\noindent
which still features a large, but positive, Kitaev term. Since  first-principle considerations indicated a negative $K$, we used the parametrization given by Eq.~(\ref{eq_set}).

% ============================================================================
{\it SWT details.}---%
% ============================================================================
Using SWT (see SM~\cite{SM}), one can obtain an explicit expression for the magnon energies at the M$_1$ point in the polarized phase for the two principal field directions, $E_{{\rm M_1},b}$ and $E_{{\rm M_1},a^*}$, with the difference of their squares for $S=1/2$ given by 
\vskip -0.05cm
\noindent
\begin{align}
\label{eq_EM_Ediff}
&E_{{\rm M_1},b}^2-E_{{\rm M_1},a^*}^2\\
&={\sf J}_{z\pm}^2+2{\sf J}_{\pm\pm}\big(2g\mu_B B +{\sf J}_1(\Delta_1-3)-3{\sf J}_3(1+\Delta_3) \big), \nonumber
\end{align}
\vskip -0.05cm
\noindent
where we neglected small ${\sf J}_2$ and assumed isotropic in-plane g-factor, $g_{a^*}\! \approx\! g_b$. As one can see, $\Delta E_{\rm M}\!\equiv \! E_{{\rm M_1},b} -E_{{\rm M_1},a^*}$ can be non-zero only because  of the bond-dependent exchanges ${\sf J}_{z\pm}$ or ${\sf J}_{\pm\pm}$. As we argue in SM~\cite{SM}, large  ${\sf J}_{\pm\pm}$ is  incompatible with the INS data, making significant  ${\sf J}_{z\pm}$ essential to explain the large $\Delta E_{\rm M}$.

\begin{figure}[t]
\includegraphics[width=\linewidth]{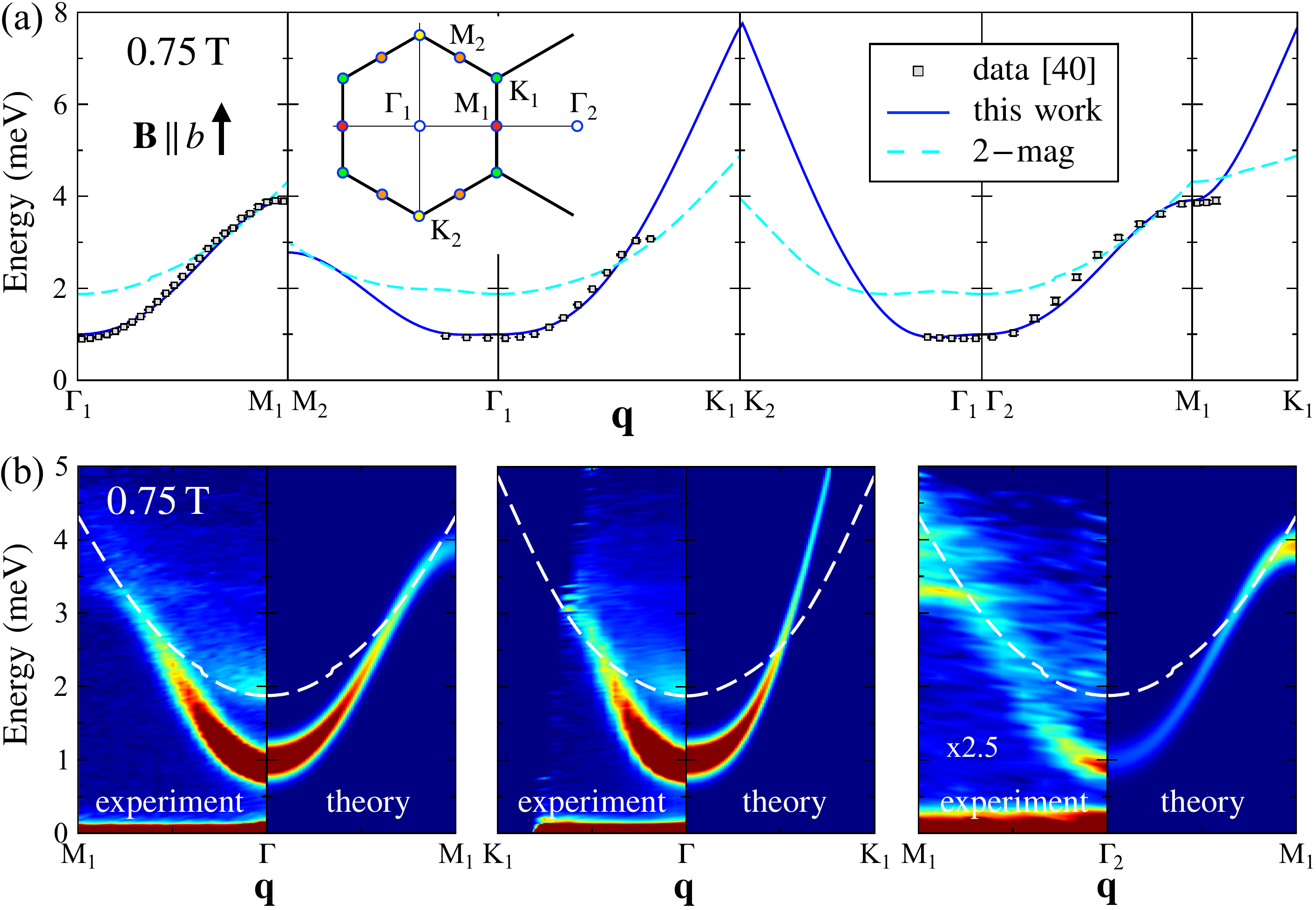}
\vskip -0.2cm
\caption{(a) and (b) Same as in Figs.~\ref{fig_spectrum}(a) and \ref{fig_spectrum}(b), respectively, for $B\!=\!0.75$~T.}
\label{FigEM:H_075T_INS}
\vskip -0.5cm
\end{figure}

% ============================================================================
{\it More INS fits.}---%
% ============================================================================
In addition to the $B\!=\!3$~T data fits in Figs.~\ref{fig_spectrum}(a) and \ref{fig_spectrum}(b), we provide further comparison of our theory with the INS measurements from Ref.~\cite{Broholm_BCAO} for $B\!=\!0.75$~T  in Fig.~\ref{FigEM:H_075T_INS}, yielding additional support for the proposed parameter set \eqref{eq_set_Jpp} for BaCAO.

The theoretical results for the dynamical structure factor presented in Figs.~\ref{fig_spectrum}(b) and \ref{FigEM:H_075T_INS}(b) are obtained from
\vskip -0.05cm
\noindent
\begin{align}
\mathcal{S}(\mathbf{q},\omega)=\sum_{\alpha,\beta}\left(\delta_{\alpha \beta}-\frac{q_\alpha q_\beta}{q^2}\right)\mathcal{S}^{\alpha \beta}(\mathbf{q},\omega),
\end{align}
\vskip -0.05cm
\noindent
with the dynamical spin correlation function 
\vskip -0.1cm
\noindent
\begin{align}
\mathcal{S}^{\alpha \beta}(\mathbf{q},\omega)=\frac{1}{\pi} \text{Im} \int_{-\infty}^\infty dt  e^{i\omega t} \  i\langle \mathcal{T} S^\alpha_\mathbf{q}(t) S^\beta_{-\mathbf{q}}(0)\rangle,
\end{align}
\vskip -0.10cm
\noindent
which is calculated using SpinW~\cite{Toth_2015} with a Gaussian broadening of 0.4~meV to match the width of the experimental peaks. Since a quantitative fit of the neutron scattering intensity is out of the scope of this paper, we assumed a fully isotropic g-factor.

% ==============================================================================
% ==============================================================================

% ============================================================================
{\it DMRG details.}---%
% ============================================================================
DMRG calculations were performed on $L_x\!\times\!L_y$-site honeycomb-lattice cylinders of width $L_y\!=\!12$  (6 honeycomb cells), using the ITensor library~\cite{itensor}. The results were obtained on the so-called XC-cylinders~\cite{j1j2-steve1}, in which one of the nearest-neighbor bonds is along the $x$-direction. We performed a sufficient number of DMRG sweeps to reach a maximum bond dimension of $m\!=\!1600$ and to ensure good convergence with a truncation error of $\mathcal{O}(10^{-5})$. 
The spin Hamiltonian does not possess any symmetry that can be utilized in DMRG, and the ground states are allowed to spontaneously break the remaining lattice symmetries, mimicking the thermodynamic limit in 2D~\cite{tt'j} and enabling us to measure the local orders directly. 

We employ  both ``scans'' (with a Hamiltonian parameter varied along the $x$-axis) and  ``non-scans'' (all parameters fixed). Their combination has been successfully used  in various other models and lattices~\cite{Zhu2015J1J2,Zhu2017YMGO,Zhu2018Topography,Zhu2013J1J2XYHoneycomb, shengtao_j1j3,nematic2023,aRu_saga}. 

For the non-scans, we used  $12\!\times\!12$ cylinders with an aspect ratio that has been demonstrated to closely approximate the 2D  thermodynamic limit~\cite{FS}.  Fig.~\ref{fig_spectrum}(d) shows an example of such a non-scan for the proposed BaCAO set (\ref{eq_set_Jpp}). 

\begin{figure}[t]
\includegraphics[width=\linewidth]{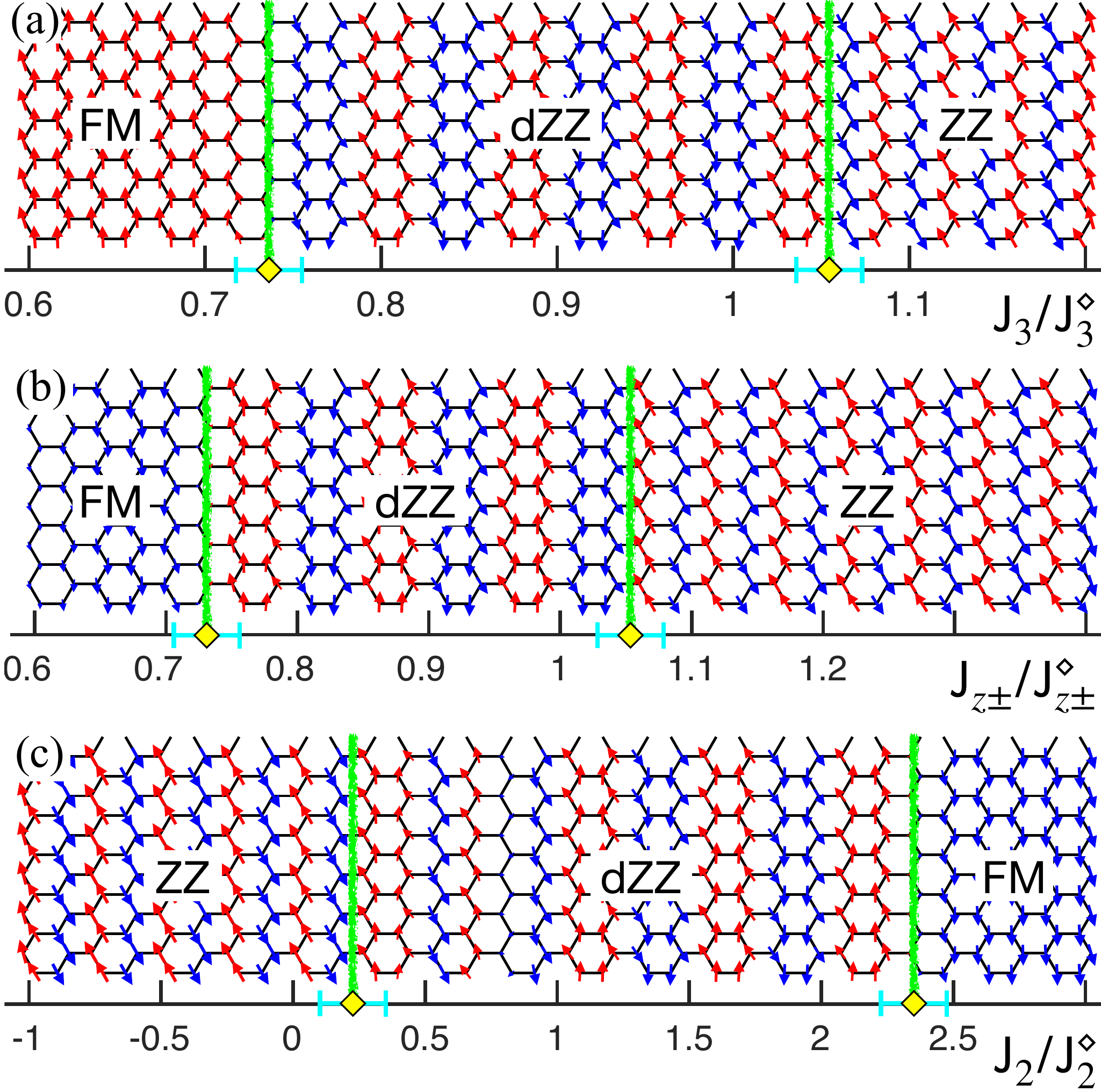}
\vskip -0.2cm
\caption{DMRG scans  along  (a) ${\sf J}_3/{\sf J}_3^\diamond$, (b)  ${\sf J}_{z\pm}/{\sf J}_{z\pm}^\diamond$, and (c) ${\sf J}_2/{\sf J}_2^\diamond$ cuts through the reference set (\ref{eq_set8_Jpp}), shown as white diamond in Fig.~\ref{fig_spectrum}(c). Phase boundaries of the dZZ region with FM and ZZ phases are marked by vertical lines and yellow diamonds. Spins are  in the $yz$ plane as in Fig.~\ref{fig_spectrum}(d).}
\label{FigEM:Scans}
\vskip -0.5cm
\end{figure}

For the scans, we  used  longer cylinders with $L_x\!=\!32$ and varied ${\sf J}_3$, ${\sf J}_{z\pm}$, or ${\sf J}_2$ parameters along their  length. The representative scans shown  in Figs.~\ref{FigEM:Scans}(a)-(c) demonstrate the exploratory power of this approach. The parameter set anchoring these scans is marked by an empty diamond in Fig.~\ref{fig_spectrum}(c); it is slightly shifted from the set proposed in Eq.~(\ref{eq_set_Jpp}) (red diamond), and is given by 
\vskip -0.05cm
\noindent
\begin{align}
\{{\sf J}_1, \Delta_1, {\sf J}_{\pm\pm}, {\sf J}_{z\pm}\}^\diamond=\{-6.64, 0.35,0.09, -3.56\}
\label{eq_set8_Jpp}
\end{align}
\vskip -0.15cm
\noindent
and $\{{\sf J}_2,{\sf J}_3,\Delta_3\}^\diamond\!=\!\{-0.2, 1.78,0.007\}$, all in meV, except for $\Delta_n$. We refer to the parameter values of this set as  a reference set   $\{{\sf J}^\diamond_{z\pm}, {\sf J}^\diamond_3,{\sf J}^\diamond_2\}\!=\!\{-3.56,1.78,-0.2\}$~meV in Figs.~\ref{FigEM:Scans}(a)-(c) and below.

The scans in Figs.~\ref{FigEM:Scans}(a) and \ref{FigEM:Scans}(b)  demonstrate the extent of the dZZ phase along the 1D  ${\sf J}_3$ and  ${\sf J}_{z\pm}$ cuts and allow us to determine phase boundaries; the scans are shown in Fig.~\ref{fig_spectrum}(c) as blue arrows. Fig.~\ref{FigEM:Scans}(c)  explores the stability of the dZZ phase in the ${\sf J}_2$ direction.  Individual non-scans along these scans  were also performed to verify the extent and  nature of the intermediate phases. 

The ${\sf J}_3$-scans of the same type were also performed for several values of ${\sf J}_{z\pm}/{\sf J}^\diamond_{z\pm}\!=\!0,0.25,0.5,1.5,$ and 2.5, allowing us to extrapolate  the 1D phase boundaries of the intermediate phase between FM and ZZ in Fig.~\ref{fig_spectrum}(c) to complete the 2D phase diagram.   As discussed in the main text, away from the region relevant to BaCAO, the triple-zigzag state appears instead of the dZZ for lower ${\sf J}_{z\pm}$, while for larger ${\sf J}_{z\pm}$, a state  resembling an SL occurs. At ${\sf J}_{z\pm}\!=\!0$, the intermediate phase is very narrow, in agreement with Ref.~\cite{shengtao_j1j3}.

Similar explorations using DMRG scans vs field for the parameter set in  Eq.~(\ref{eq_set_Jpp}) are discussed in SM~\cite{SM}.

% ==============================================================================
% ==============================================================================

% ============================================================================
{\it Other aspects.}---%
% ============================================================================
The magnetoelastic coupling's role in affecting BaCAO's spin model remains an open question~\cite{LP90,Tsirlin_BCAO_2024,Ong_BCAO_kappa_2024}. While the field-induced first-order transitions may alter the model, given how subtle some of the properties are---the dZZ state, the out-of-plane tilt, and the UUD phase---, which are all successfully described by our model, we find it to be an unlikely scenario.

%\newpage 

%\end{document}

% ==============================================================================

%---------------------------------------------------------------------------
\newpage 
%\ \
%\newpage 
%\ \
%\newpage
\onecolumngrid
%---------------------------------------------------------------------------
\begin{center}
{\large\bf BaCo$_2$(AsO$_4$)$_2$: Strong Kitaev, After All: Supplemental Material}\\ 
\vskip 0.35cm
Pavel A. Maksimov,$^{1,2}$ \begin{CJK*}{UTF8}{}
Shengtao Jiang (\CJKfamily{gbsn}蒋晟韬),$^{3}$
\end{CJK*}
L. P. Regnault,$^{4,5}$ and A. L. Chernyshev$^6$\\
\vskip 0.15cm
{\it \small $^1$Bogolyubov Laboratory of Theoretical Physics, Joint Institute for Nuclear Research, Dubna, Moscow region 141980, Russia}\\
{\it \small $^2$M. N. Mikheev Institute of Metal Physics of Ural Branch of Russian Academy of Sciences, \\ S. Kovalevskaya St. 18, 620990 Yekaterinburg, Russia}\\
{\it \small $^3$Stanford Institute for Materials and Energy Sciences, \\ SLAC National Accelerator Laboratory and Stanford University, Menlo Park, California 94025, USA}\\
{\it \small $^4$Institut Laue Langevin, 71 avenue des Martyrs, CS 20156, 38042 Grenoble Cedex 9, France}\\
{\it \small $^5$Laboratoire de Magn{\'e}tisme et Diffraction Neutronique, \\ CEA-Grenoble, 17 rue des Martyrs, 38054 Grenoble Cedex 9, France}\\
{\it \small $^6$Department of Physics and Astronomy, University of California, Irvine, California 92697, USA}\\
{\small (Dated: \today)}\\
%\vskip -0.5cm \
\end{center}
\vskip 0.2cm 
%---------------------------------------------------------------------------
\twocolumngrid
%\end{widetext}
%---------------------------------------------------------------------------
% ==============================================================================
\setcounter{page}{1}
\thispagestyle{empty}
% ==============================================================================
\makeatletter
\renewcommand{\c@secnumdepth}{0}
\makeatother
%%%%%%%%%%%%%%%%%%%%%%%%%%%%%%%%%%%%%%%%%%%%%%%%%%
%\renewcommand{\thesection}{\Roman{section}}
%%%%%%%%%%%%%%%%%%%%%%%%%%%%%%%%%%%%%%%%%%%%%%%%%%
\setcounter{section}{0}
% ==============================================================================
%%%%%%%%%%%%%%%%%%%%%%%%%%%%%%%%%%%%%%%%%%%%%%%%%%
\renewcommand{\theequation}{S\arabic{equation}}
%%%%%%%%%%%%%%%%%%%%%%%%%%%%%%%%%%%%%%%%%%%%%%%%%%
\setcounter{equation}{0}
% ==============================================================================
%%%%%%%%%%%%%%%%%%%%%%%%%%%%%%%%%%%%%%%%%%%%%%%%%%
\renewcommand{\thefigure}{S\arabic{figure}}
%%%%%%%%%%%%%%%%%%%%%%%%%%%%%%%%%%%%%%%%%%%%%%%%%%
\setcounter{figure}{0}
% ==============================================================================
%%%%%%%%%%%%%%%%%%%%%%%%%%%%%%%%%%%%%%%%%%%%%%%%%%
\renewcommand{\thetable}{S.\Roman{table}}
%%%%%%%%%%%%%%%%%%%%%%%%%%%%%%%%%%%%%%%%%%%%%%%%%%
\setcounter{table}{0}
% ==============================================================================
%\vspace{-0.3cm}

%\vspace{-0.2cm}
% ==============================================================================
\section{Experimental details}
% ==============================================================================
\vskip -0.2cm
% ==============================================================================
\subsection{INS experiments at IN22/ILL}
% ==============================================================================
\vskip -0.2cm

The BaCAO single crystal, grown by a flux method and shaped as a dark-purple platelet of dimensions $14\!\times\!14\!\times\!1$~mm$^{3}$, was the same as the one used in previous neutron-scattering experiments~\cite{LP18}. Within the hexagonal-cell description, the lattice parameters at $T\!\approx\! 2$ K were $a\!=\!b \approx 4.95$~{\AA} and $c \approx 23.13$~{\AA}, as determined from elastic scans across the structural Bragg reflections (3,0,0) and (0,0,9), respectively.  

The INS experiments were carried out on the TAS IN22, a high-flux instrument with polarized-neutron capabilities, installed at the end position of the thermal H25, m=2 supermirror guide at the Institut Laue Langevin (ILL), Grenoble. We used a vertical-focusing/flat-horizontal pyrolitic graphite (PG-002) monochromator and a flat-vertical/horizontal-focusing PG-002 analyzer, at several fixed final neutron wave vectors, $k_{f}\!=\!1.64$ \AA $^{-1}$, $1.97$ \AA$^{-1}$, and $2.662$ \AA$^{-1}$. The elastic resolution in energy, measured on a vanadium incoherent scatterer, was typically $0.35$, $0.5$, and $0.9$ meV, respectively. In all cases, a 7-cm thick PG filter was placed in the scattered beam, in order to avoid parasitic higher-order neutrons from being scattered by the analyzer and detected (especially those at $2k_{f}$).

% ==============================================================================
\begin{figure}[t]
\centering
\includegraphics[width=8cm]{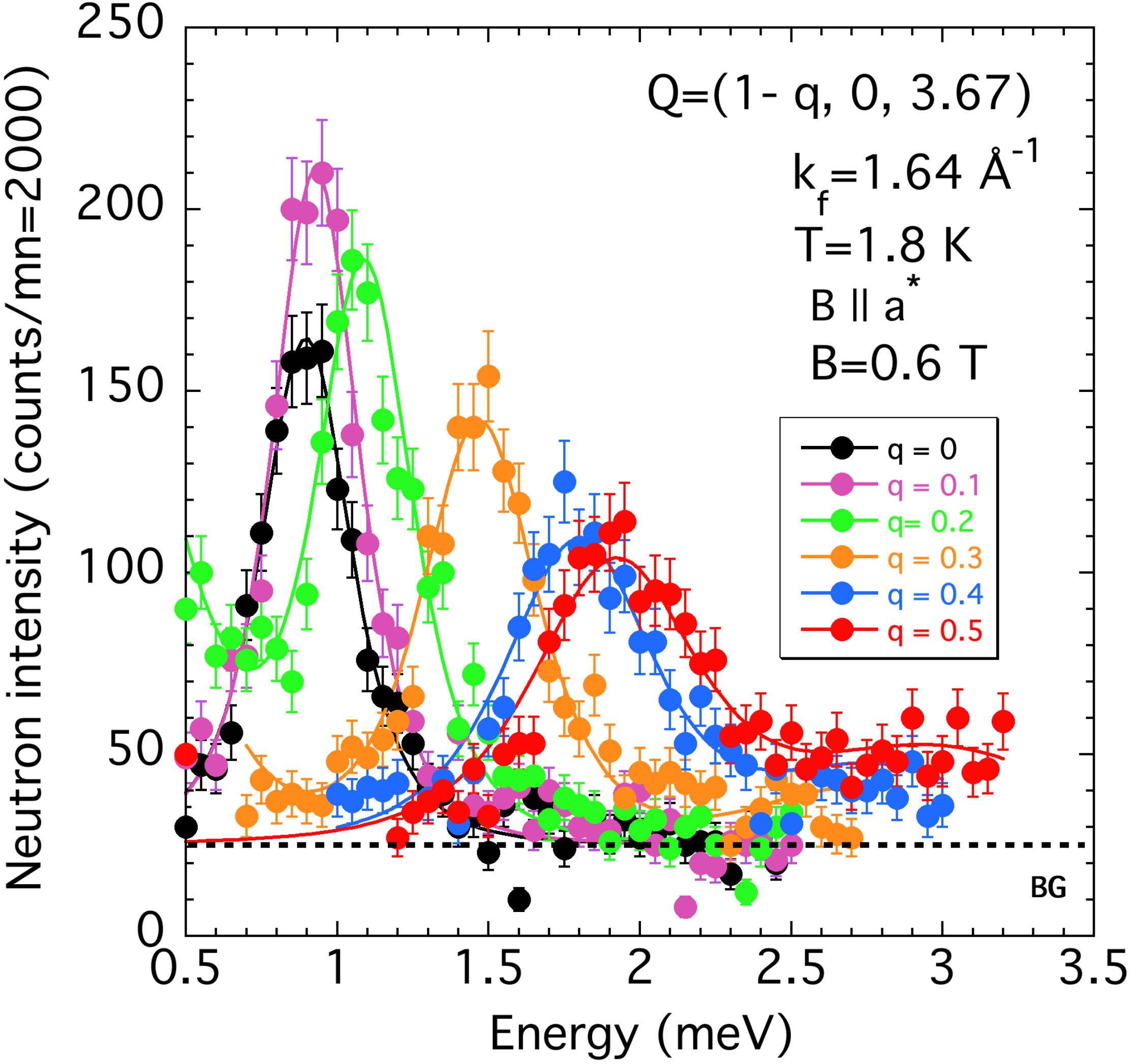}
\vskip -0.2cm
\caption{Constant-Q scans at various scattering vectors ${\bf Q}\!=\!(1 - q,0,3.67)$ and fixed $k_f\!=\!1.64~\AA^{-1}$, in a magnetic field of $0.6$~T applied parallel to ${\bf a^*}$. The solid lines are fits to Voigt functions as described in the text. The dashed line (BG) represents the spectrometer's neutron background.}
\label{SM:scan06}
\vskip -0.5cm
\end{figure}
% ==============================================================================

For the experiments in a magnetic field ${\bf B}\!\parallel\! {\bf a^*}$,  the sample was mounted on the cold finger of a rotating sample-holder insert in the ILL 3.8-T horizontal-field cryomagnet and kept at $T\!=\!1.8\text{~K}$. The magnetic field was first aligned parallel to ${\bf a^*}$ within an accuracy of $\pm 1^\circ$, by means of the rotating sample-holder stick, and kept fixed in direction during the various scans by rotating  the cryomagnet adequately.

Complementary measurements were taken in a vertical field parallel to ${\bf b}$, by using the ILL 5-T vertical-field cryomagnet.

In all cases, the sample was aligned with the ${\bf b}$ crystallographic axis vertical. This orientation allowed the survey of scattering vectors ${\bf Q} \!=\! (Q_{a},0,Q_{c}$), also offering the possibility to check the temperature and field dependencies of magnetic Bragg peaks associated with the various field-induced magnetic phases, all characterized by propagation vectors ${\bf k}\!=\!(k_a,0,k_c)$~\cite{LP18,LP84,LP90,LP79}.

\begin{figure}[t]
\centering
\includegraphics[width=8cm]{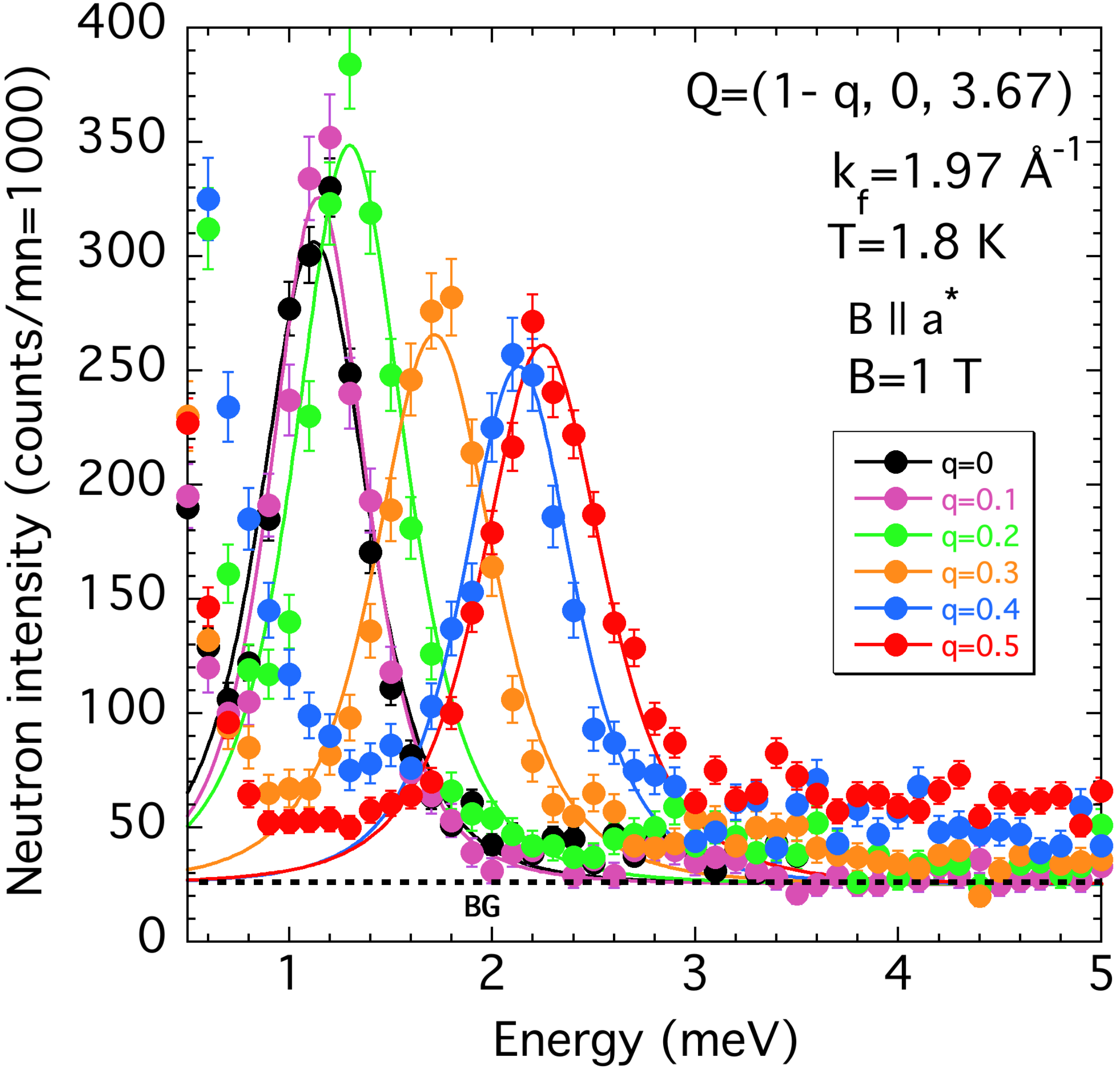}
\vskip -0.2cm
\caption{Same as in Fig.~\ref{SM:scan06} for magnetic field 1~T.}
\label{SM:scan1}
\vskip -0.5cm
\end{figure}

\begin{figure}[b]
\vskip -0.5cm
\centering
\includegraphics[width=8cm]{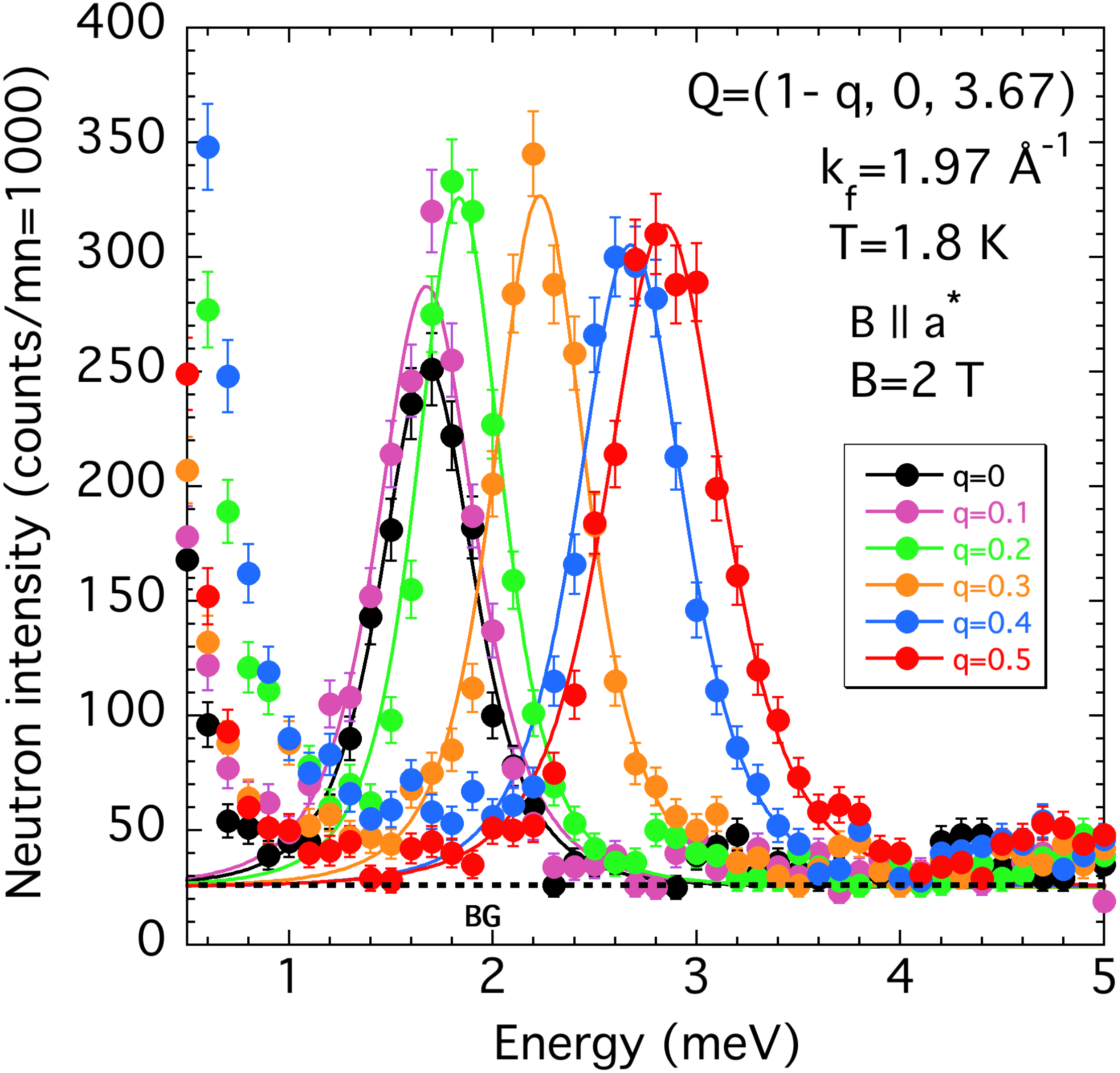}
\vskip -0.2cm
\caption{Same as in Fig.~\ref{SM:scan06} for magnetic field 2~T.}
\label{SM:scan2}
\end{figure}

\begin{figure}[t]
\centering
\includegraphics[width=8cm]{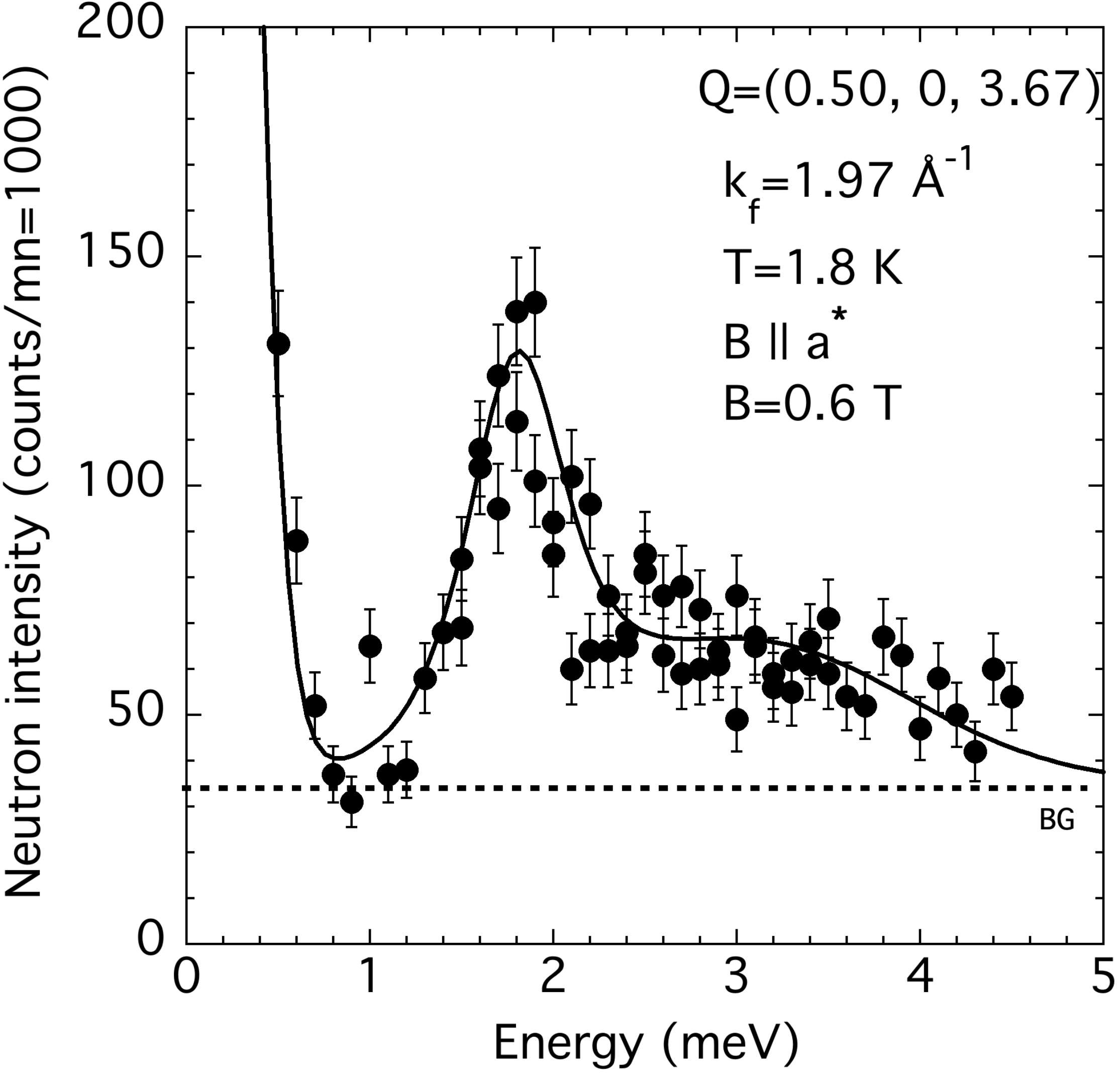}
\vskip -0.2cm
\caption{Constant-Q scan at the scattering vector ${\bf Q}\!=\!(0.5, 0,3.67)$ and fixed $k_f\!=\!1.97~\AA^{-1}$, for a magnetic field applied parallel to ${\bf a^*}$.}
\label{Fig-2}
\vskip -0.5cm
\end{figure}

The dispersions of magnetic excitations along the ${\bf a^*}$ direction were measured at scattering vectors ${\bf Q} \!=\! (Q_a,0,3.67)$, with the $Q_a$ component ranging from $1$ to $0.5$, for magnetic-field amplitudes up to $3$ T. Typical energy scans performed at magnetic fields of $0.6$~T, 1~T, and 2~T applied along ${\bf a^*}$, are shown in Figs.~\ref{SM:scan06}--\ref{SM:scan2} ($k_f \!=\! 1.64$ \AA$^{-1}$) and Fig.~\ref{Fig-2} ($k_f \!=\! 1.97$ \AA$^{-1}$). The magnetic-excitation energies have been extracted by fitting the experimental data to Voigt functions, which correspond to the convolution of a Gaussian function (the instrument resolution function) with a Lorentzian function (the physical function).

The INS energy scan data presented in Figs.~\ref{SM:scan06}--\ref{Fig-2} and in Fig.~\ref{Fig-4} can be accessed directly at Zenodo open data repository~\cite{Zenodo_data}.

% ==============================================================================
% ==============================================================================

% ==============================================================================
% ==============================================================================

\vspace{-0.3cm}
% ==============================================================================
\subsection{INS experiments at DN1/Siloe}
% ==============================================================================
\vskip -0.2cm

The earlier experiments, reported in Refs.~\cite{LP84,LP90}, were carried out at a fixed incident neutron wavelength of $\lambda_{i} \!=\! 2.4$~\AA, on the thermal-neutron TAS DN1, installed on a radial beam tube of the Silo{\'e} medium-flux $35$-MW reactor at CEA-Grenoble. For these INS measurements, we used flat-vertical/flat-horizontal PG-002 monochromator and analyzer devices. A 5-cm thick PG filter was placed in the incident beam, in order to remove neutrons with wavelengths $\lambda_{i}/2$, $\lambda_{i}/3$, $\dots$, scattered by the monochromator prior to reaching the sample. In this configuration, a typical elastic energy resolution of about $0.8$ meV could be achieved.

The single crystal, shaped as a platelet of dimensions $15\!\times\!10\!\times\!1$~mm$^3$, was aligned with the ${\bf b}$ axis vertical and mounted in the CEA-Grenoble 10-T vertical-field cryomagnet and kept at $T\!=\!4.2\text{~K}$.

The dispersions of magnetic excitations along the ${\bf a^*}$ direction were investigated by performing constant-$Q$ scans at scattering vectors ${\bf Q} \!=\! (q_a,0,5)$, with the $q_a$ component varying from $0$ to $0.5$~r.l.u., at temperatures below $2$ K and magnetic-field values ranging from 0 up to $6$ T~\cite{LP84,LP90}. Examples of magnon dispersion curves for various magnetic fields parallel to ${\bf b} $ are shown in Fig.~\ref{fig_INS} in the main text~\cite{LP84,LP90}.

% ==============================================================================
% ==============================================================================

\vspace{-0.3cm}
% ==============================================================================
\section{Spin-wave theory details}
% ==============================================================================
\vskip -0.2cm
% ==============================================================================
\subsection{LSWT in the polarized state}
% ==============================================================================
\label{Sec_LSWT}
\vskip -0.2cm

In this work, we study the extended Kitaev-Heisenberg--${\sf J}_2$--${\sf J}_3$ model. Written within the crystallographic reference frame $\{x,y,z\}$, with $\hat{\bf x}\!\parallel\! {\bf a^*}$, $\hat{\bf y}\!\parallel\! {\bf b}$, and $\hat{\bf z}\!=\!\hat{\bf x} \times \hat{\bf y}$, it is given by Eqs.~(\ref{eq_HXXZ})+(\ref{eq_HJpm}) in the main text, in which further-neighbor interactions are kept purely $XXZ$ with no bond-dependent components in their  exchange matrices, and ${\sf J}_2$ is also purely $XY$ ($\Delta_2\!=\!0$). The nearest-neighbor exchange matrix contains all four symmetry-allowed terms:  ${\sf J}_1,\Delta_1,{\sf J}_{\pm \pm}$, and ${\sf J}_{z\pm}$. 

In the field-polarized state, the spins' local reference frame is along the field. For the linear SWT (LSWT), one performs an axes rotation to align the $z$ axis with either ${\bf B}\!\parallel\! b$ or ${\bf B}\!\parallel\! a^*$, and performs the lowest-order Holstein-Primakoff transformation for the spin-$S$ model
\begin{align}
&S^+_{i,\alpha}\approx \sqrt{2S} a_{i,\alpha}, \ \ \ \ 
S^z_{i,\alpha}=S-a_{i,\alpha}^\dagger a^{\phantom \dagger}_{i,\alpha},
\end{align}
where $a_{i,1}\equiv a_i$, $a_{i,2}\equiv b_i$ for the two sublattices of the honeycomb lattice. The  Fourier transformation
\begin{align}
a^{\phantom{\dag}}_{i,\alpha}=
\frac{1}{\sqrt{N}} \sum_{{\bf q}} e^{i{\bf q}({\bf r}_i+{\bm \rho}_\alpha)}
a^{\phantom{\dag}}_{\alpha {\bf q}},
\label{eq_fourier}
\end{align}
with $N$ being the number of  unit cells, ${\bf r}_i$ and  ${\bm \rho}_\alpha$ being the unit cell and sublattice coordinates, respectively, with  ${\bf q}$ summing over  the Brillouin zone of the honeycomb lattice, yields the LSWT Hamiltonian that can be written in a matrix form as
\begin{align}
\mathcal{H} =\frac{1}{2}\sum_\mathbf{q}\hat{\mathbf{x}}_\mathbf{q}^\dagger \mathbf{H_q}\hat{\mathbf{x}}_\mathbf{q}^{\phantom{\dagger}},
\label{eq_hamak}
\end{align}
where $\hat{\mathbf{x}}_\mathbf{q}=\big( a^{\phantom{\dagger}}_{\bf q},b^{\phantom{\dagger}}_{\bf q},a^\dagger_{\bf -q},b^\dagger_{\bf -q}\big)$. 

The general form of the Hamiltonian matrix $\mathbf{H_q}$ in the polarized phase for both field directions is given by
\begin{eqnarray}
\mathbf{H_q}=
\left( \begin{array}{cccc} 
A_{\bf q} &  B_{\bf q} & F_{\bf q} & C_{\bf q}\\ 
B^*_{\bf q} &  A_{\bf q} & C_{\bf -q} & F_{\bf q}\\
F_{\bf q} &  C^*_{\bf -q} & A_{\bf q} & B_{\bf q}\\
C^*_{\bf q} &  F_{\bf q} & B^*_{\bf q} & A_{\bf q}
\end{array}\right).
\label{eq_hammatrix}
\end{eqnarray}
The elements of the matrix for ${\bf B} \parallel b$ are given by
\begin{align}
A_{\bf q}&=g_y\mu_B B-3S({\sf J}_1+{\sf J}_3)+3S(({\sf J}_2 + {\sf J}_{2z})\gamma_{2\bf q} - 2{\sf J}_2),\nonumber\\
B_{\bf q}&=\frac{3S}{2} \left[ ({\sf J}_1+{\sf J}_{1z}) \gamma_{\bf q} + ({\sf J}_3 +{\sf J}_{3z})\gamma_{3\bf q}
%\right.\nonumber\\&\left.
+2{\sf J}_{\pm\pm}\gamma'_{\bf q}\right],\nonumber\\
C_{\bf q}&=\frac{3S}{2} \left[ ({\sf J}_1-{\sf J}_{1z}) \gamma_{\bf q}+ ({\sf J}_3-{\sf J}_{3z}) \gamma_{3\bf q}+2{\sf J}_{\pm\pm}\gamma'_{\bf q}\right.\nonumber\\
\label{SM_eq_ABCF_b}
&\left.-2i{\sf J}_{z\pm}\gamma''_{\bf q}\right],\\ 
F_{\bf q}&=3S({\sf J}_2-{\sf J}_{2z}) \gamma_{2\bf q},\nonumber
\end{align}
and, respectively, for ${\bf B} \parallel a^*$
\begin{align}
A_{\bf q}&=g_x\mu_B B -3S({\sf J}_1+{\sf J}_3)+3S(({\sf J}_2 + {\sf J}_{2z})\gamma_{2\bf q} - 2{\sf J}_2),\nonumber\\
B_{\bf q}&=\frac{3S}{2} \left[ ({\sf J}_1+{\sf J}_{1z}) \gamma_{\bf q} + ({\sf J}_3 +{\sf J}_{3z})\gamma_{3\bf q}
-2{\sf J}_{\pm\pm}\gamma'_{\bf q}\right],\nonumber\\
C_{\bf q}&=\frac{3S}{2} \left[ ({\sf J}_{1z}-{\sf J}_1) \gamma_{\bf q} +({\sf J}_{3z}-{\sf J}_3) \gamma_{3\bf q}+2{\sf J}_{\pm\pm}\gamma'_{\bf q}\right.\nonumber\\
\label{SM_eq_ABCF_a*}
&\left.-2i{\sf J}_{z\pm}\gamma'_{\bf q}\right],\\ 
F_{\bf q}&=3S({\sf J}_2-{\sf J}_{2z}) \gamma_{2\bf q},\nonumber
\end{align}
where we define
\begin{align}
{\sf J}_{nz}=\Delta_n {\sf J}_n,
\end{align}
and the hopping amplitudes are
\begin{eqnarray}
\gamma_{\bf q}&&=\frac{1}{3}\sum_{\alpha=1}^3 e^{i{\bf q}{\bm \delta}^{(1)}_\alpha}, \\ 
\gamma^{\prime}_{\bf q}&&=\frac{1}{3}\sum_{\alpha=1}^3 \cos\tilde{\varphi}_\alpha e^{i{\bf q}{\bm \delta}^{(1)}_\alpha},\ 
\gamma^{\prime\prime}_{\bf q}=\frac{1}{3}\sum_{\alpha=1}^3 \sin\tilde{\varphi}_\alpha e^{i{\bf q}{\bm \delta}^{(1)}_\alpha},
\nonumber\\
\gamma_{2\bf q}&&=\frac{1}{6}\sum_{\alpha=1}^6 e^{i{\bf q}{\bm \delta}^{(2)}_\alpha},\ \
\gamma_{3{\bf q}}=\frac{1}{3}\sum_{\alpha=1}^3 e^{i{\bf q}{\bm \delta}^{(3)}_\alpha},\nonumber
\end{eqnarray}
where $\tilde{\varphi}_\alpha\!=\!\{0,2\pi/3,-2\pi/3\}$ are the bond angles with the $x$ axis in Fig.~\ref{fig_lattice} as before and ${\bm \delta}^{(n)}_\alpha$ are the $n$th-neighbor vectors.
 
The two magnon branches of the LSWT spectrum $E^{(\alpha)}_{\bf q}$ are obtained  as the eigenvalues  of $\mathbf{g}\mathbf{H_q}$, where $\mathbf{g}$ is a diagonal matrix $\{1,1,-1,-1\}$, using the standard diagonalization procedure~\cite{Colpa}.

\vspace{-0.2cm}
% ==============================================================================
\subsection{Energies at the ${\rm M_1}$ point}
% ==============================================================================
\vskip -0.2cm

For ${\bf B} \!\parallel\! b$, the magnon energies  along the $\Gamma \text{M}_1$ direction are independent of ${\sf J}_{z\pm}$, because its hopping amplitude $\gamma''_{\bf q}$ in Eq.~(\ref{SM_eq_ABCF_b}) is $\equiv\!0$ for ${\bf q}\! \parallel\! a^*$, making  the $\Gamma \text{M}_1$ INS spectra  blind to ${\sf J}_{z\pm}$. For ${\bf B} \!\parallel\! a^*$, the $\Gamma \text{M}_1$ energies do depend on ${\sf J}_{z\pm}$ as the hopping amplitude of the corresponding term in the LSWT matrix (\ref{SM_eq_ABCF_a*}) is non-zero. 

Using (\ref{SM_eq_ABCF_a*}) and (\ref{SM_eq_ABCF_b}), one can obtain the  difference between the energies for the $b$ and $a^*$ field directions at the ${\bf q}\!=\!\text{M}_1$ point for the lower magnon branch  
\begin{align}
\label{SM_eq_dE}
&\frac{E_{{\rm M_1},b}^2-E_{{\rm M_1},a^*}^2}{4S^2}\\
&={\sf J}_{z\pm}^2+2{\sf J}_{\pm\pm}\left(\frac{g\mu_B B}{S} +{\sf J}_1(\Delta_1-3)-3{\sf J}_3(1+\Delta_3) \right),
\nonumber
\end{align}
where we  have neglected ${\sf J}_2$.

One can see from (\ref{SM_eq_dE}) that in order to explain the observed large energy difference at the M$_1$ point in Fig.~\ref{fig_lattice}, a large ${\sf J}_{z\pm}\!\sim\!{\sf J}_1$ is needed, such as the one proposed for BaCAO in our set \eqref{eq_set_Jpp}. 

Theoretically, a less significant, but still substantial ${\sf J}_{\pm\pm}$ alone can also fit such an energy difference for the field value of 3~T in Fig.~\ref{fig_lattice}. The corresponding parameter set without ${\sf J}_{z\pm}$, which can be found using the same procedure as  the one described below: $\{{\sf J}_1, \Delta_1, {\sf J}_{\pm\pm}, {\sf J}_{z\pm}\}^\prime=\{-6.78, 0.37, -0.41, 0.0\}$ and $\{{\sf J}_2,{\sf J}_3,\Delta_3\}^\prime\!=\!\{-0.2, 2.0, 0.17\}$, all in meV except for $\Delta_n$, can also  provide a decent fit to the rest of the INS data. However, this set:  (i) provides a field dependence of the LSWT spectra that is noticeably incompatible with the INS data at lower fields, (ii) is incapable of explaining the out-of-plane tilt of spins seen in neutron polarimetry~\cite{LP18,LP06}, and (iii) corresponds to an FM ground state according to DMRG.

\vspace{-0.2cm}
% ==============================================================================
\subsection{Earlier $XXZ$ parameter set}
% ==============================================================================
\vskip -0.2cm

As mentioned in the main text, the modeling of the earlier INS data for BaCAO in the polarized phase~\cite{LP84,LP90}  yielded a pure $XXZ$  parameter set, which is close to the $XXZ$ sector of the set advocated in the present work in (\ref{eq_set_Jpp}), and is given by
\vskip -0.05cm
\noindent
\begin{align}
\{{\sf J}_1, {\sf J}_2, {\sf J}_3, \Delta_n\}=\{-6.55,-0.22, 1.72, 0.4\},
\label{eq_set_LP90}
\end{align}
\vskip -0.05cm
\noindent
with ${\sf J}_n$ in meV, and it should be noted that the exchanges in Refs.~\cite{LP84,LP90}  used an additional factor of 2 in their conventions.

\begin{figure}[t]
\includegraphics[width=\linewidth]{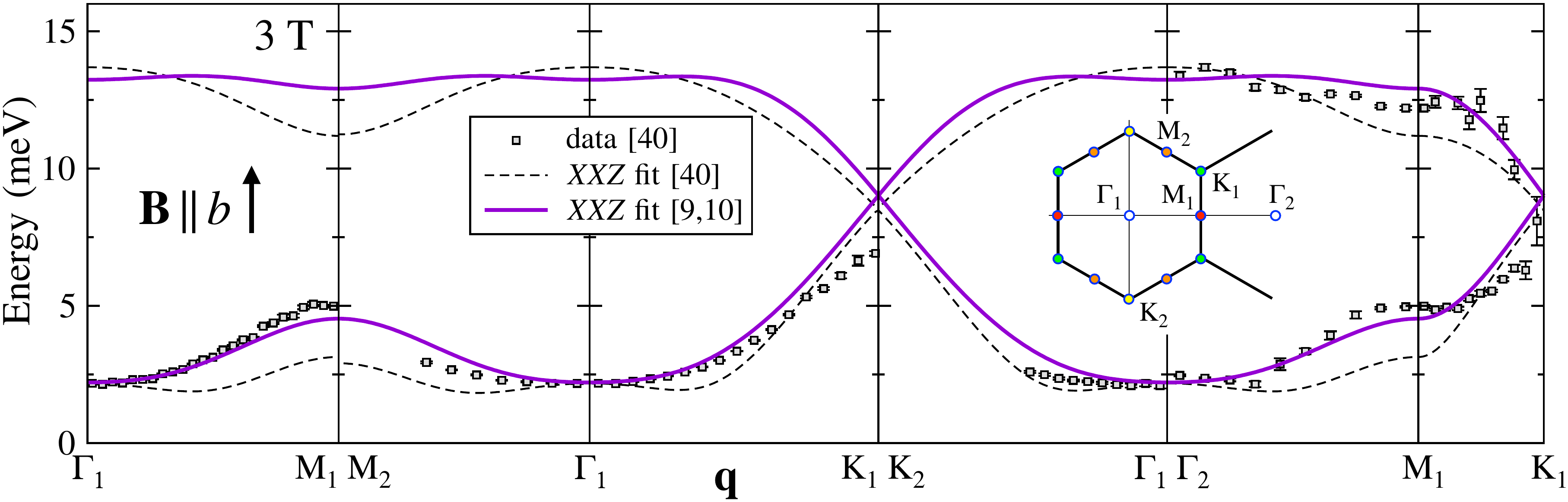}
\vskip -0.2cm
\caption{Same data from Ref.~\cite{Broholm_BCAO} for ${\bf B}\!\parallel\! b$, $B\!=\!3$~T as in Fig.~\ref{fig_spectrum}(a) with the fit using the $XXZ$ ${\sf J}_1$--${\sf J}_2$--${\sf J}_3$ model with the parameters in~(\ref{eq_set_LP90}) from Refs.~\cite{LP84,LP90} (solid line).}
\label{FigSM:LP_fit}
\vskip -0.3cm
\end{figure}

As one can see in Fig.~\ref{FigSM:LP_fit}, this parameter set provides reasonable agreement for the ${\bf B}\!\parallel\! b$, $B\!=\!3$~T data from the recent INS study~\cite{Broholm_BCAO}, especially for the $\Gamma{\rm M}_1$ and equivalent ${\bf q}$-directions. This partial success is also an indirect argument for the smallness of the ${\sf J}_{\pm\pm}$ term in the more complete model. However, similarly to the $XXZ$ model from Ref.~\cite{Broholm_BCAO}, this parameter set also fails to explain both: (i) the significant asymmetry between the ${\bf B}\!\parallel\! b$ and ${\bf B}\!\parallel\! a^*$ INS data in Figs.~\ref{fig_lattice} and \ref{fig_INS}, and (ii) the dZZ ground state and its tilt out of the plane, making the substantial bond-dependent terms of the model~\eqref{eq_HJpm}, and specifically ${\sf J}_{z\pm}$, absolutely necessary. 

% ==============================================================================
% ==============================================================================

\vspace{-0.3cm}
% ==============================================================================
\subsection{Model parameters from INS data}
% ==============================================================================
\label{Sec_INS_LSWT_parameters}
\vskip -0.2cm

One can safely assume that the polarized phase of BaCAO is free from fluctuations, a notion strongly supported by the magnetization data~\cite{Cava_2020_BaCo}. Several collections of  INS data from this phase for  representative ${\bf q}$-cuts and for the two principal in-plane field directions from Refs.~\cite{LP84,LP90,Broholm_BCAO} and from the present study, shown in Figs.~\ref{fig_lattice}, \ref{fig_INS}, \ref{fig_spectrum}, and \ref{FigEM:H_075T_INS}, are at our disposal. The minimal model (\ref{eq_HXXZ})+(\ref{eq_HJpm}) is well-justified, and the LSWT spectra can be calculated as  described above. Altogether, this seems like a recipe for a straightforward parameter-extraction exercise. 

However, there were several reasons why it turned out to be somewhat less straightforward than expected. As one can observe in the INS spectra, such as those in Fig.~\ref{FigEM:H_075T_INS}(b) and  Fig.~\ref{fig_spectrum}(b), the bottom of the two-magnon continuum can intersect the single-particle branch in the proximity of the maxima of the lowest magnon branch (K and M points) even in fields of  3~T, much higher than the critical ones, and affect the data. Specifically, the  maxima observed in Ref.~\cite{Broholm_BCAO} at the K point at 3~T and 0.75~T  are likely the bottoms of such continua, not the magnon branches. Thus, a blind ``best-fit'' LSWT  approach is likely to provide unphysical results or fail altogether if the data from the proximity of the K point  are used directly. 

The INS peaks for the upper (optical) magnon mode are wider, and the data show more scattering; see Fig.~\ref{fig_spectrum}(a) and Ref.~\cite{Broholm_BCAO}. In particular,  the optical mode near the $\Gamma$ point can be affected by its proximity to a spin-orbit exciton ($j_{eff}\!=\!3/2$) mode, which has an energy of approximately  17.5~meV~\cite{BCAO_Raman_2024,Broholm_BCAO}. Therefore,  more relaxed LSWT fitting criteria are needed for it as well. 

A minor point is a small but noticeable deviation in the 3~T INS data in Ref.~\cite{Broholm_BCAO} between the two different neutron incident energies, for instance along the $\Gamma$M$_1$ line, which also suggests using slightly less rigid criteria for the LSWT fitting.

Lastly, the LSWT Hamiltonian matrices (\ref{SM_eq_ABCF_b}) and (\ref{SM_eq_ABCF_a*}) can be used to obtain analytic expressions for the magnon energies at several high-symmetry points. This can be seen as an opportunity to reduce the number of   independent model parameters with the help of the INS data for the energies at these points and also to gain  intuition about the smallness or largeness of a subset of such parameters. However, in practice, their usefulness in quantitative fits, combined with the aforementioned uncertainties in some of the data, was limited.

All these difficulties notwithstanding, we have used the following strategy: 
(i) To extract parameters, we use only high-field, $B\!\geq\!3$~T, INS data from Refs.~\cite{LP90,Broholm_BCAO} and the present study, to eliminate the possibility of quantum fluctuations affecting our results. 
(ii) From the available INS data, we take magnon energies at the high-symmetry points ($\Gamma$, M, and K), for the two principal field directions, for three different fields, and for the two magnon modes (seven data points altogether). 
(iii) Of the seven parameters of the effective model (\ref{eq_HXXZ})+(\ref{eq_HJpm})  and the two in-plane g-factors, we fix $J_2\!=\!-0.2$~meV, using prior $XXZ$ guidance~\cite{LP84}, and $g_{a^*}\!=\!4.8$, using the field-dependence of the lowest magnon mode at $\Gamma$ from Fig.~\ref{fig_INS}(a), leaving us with seven parameters still: $\{{\sf J}_1,\Delta_1,{\sf J}_{\pm\pm},{\sf J}_{z\pm},{\sf J}_3,\Delta_3,g_b\}$. 
(iv) Lastly, we allow the 3~T LSWT energies for $\mathbf{B}\!\parallel\! b$  for the optical mode at the $\Gamma$ point, $E_2(\Gamma)$, and for the lowest mode at the K point, $E_1(\text{K}_1)$, to be adjusted upward to accommodate the effects of the exciton and the two-magnon crossing, respectively. 

This strategy has allowed us to search for  parameters that make the corresponding system of equations numerically solvable, and to achieve the best fit to the data. By following this strategy, we have found a parameter set  referred to as \#12 (for historical reasons), which yielded $g_{b}\!=\!4.85$ and  
\vskip -0.05cm
\noindent
\begin{align}
&\{{\sf J}_1, \Delta_1, {\sf J}_{\pm\pm}, {\sf J}_{z\pm}\}^{\#12}=\{-6.38, 0.37,0.24, -4.09\},\nonumber\\
\label{eq_SM_set12}
&\{{\sf J}_2,{\sf J}_3,\Delta_3\}^{\#12}\!=\!\{-0.2, 1.55, 0.09\},
\end{align}
\vskip -0.05cm
\noindent
all in meV except for $\Delta_n$, as before. The projection of this parameter set on the ${\sf J}_3$--${\sf J}_{z\pm}$ phase diagram in Fig.~\ref{fig_spectrum}(c) is shown by a green diamond. It differs from the set (\ref{eq_set_Jpp}) shown by a red diamond in the same figure by less than 10\% adjustments to the main parameters and corresponds to somewhat higher values of the Kitaev exchange.

\begin{figure}[t]
\includegraphics[width=\linewidth]{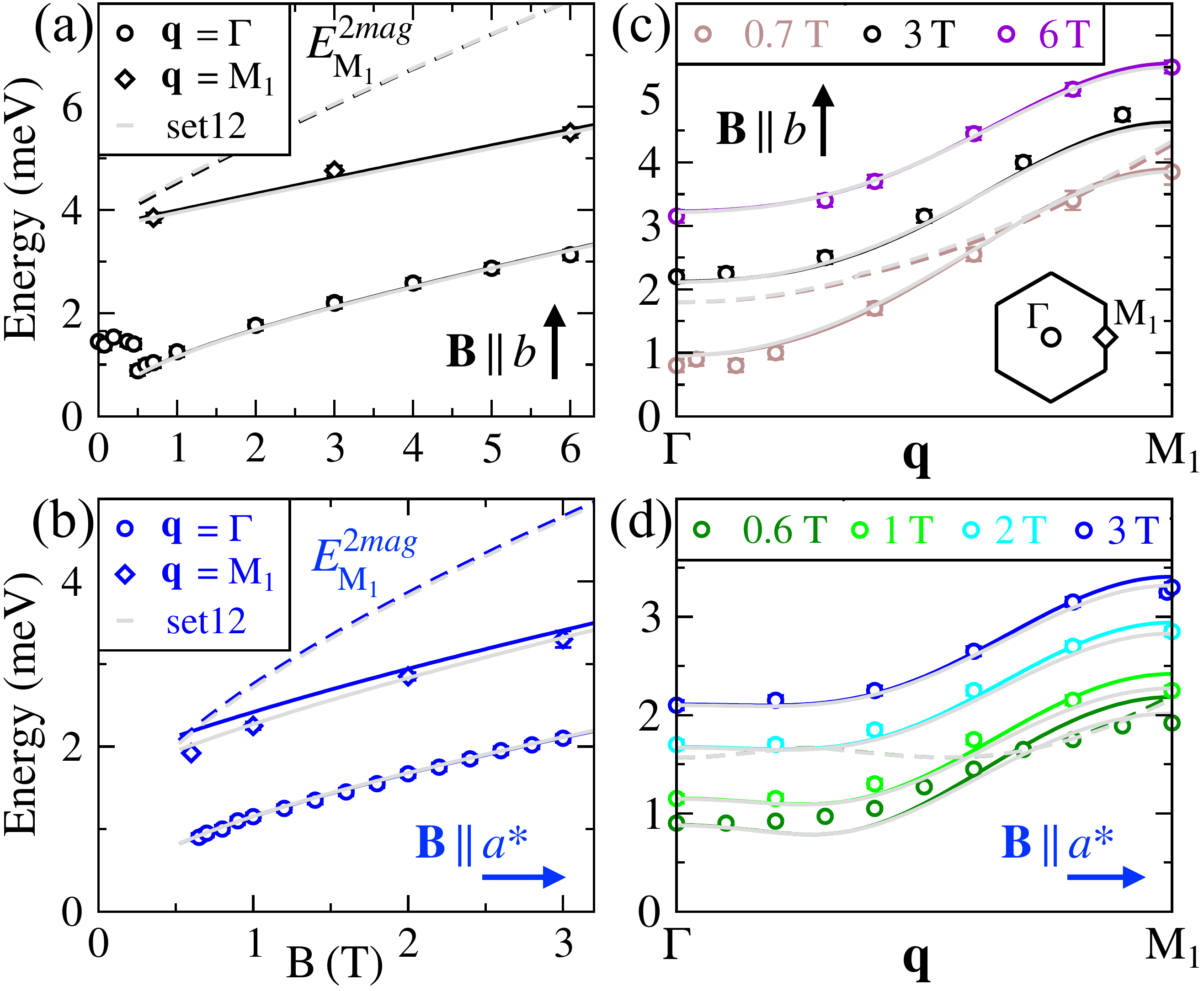}
\vskip -0.2cm
\caption{Same as in Fig.~\ref{fig_INS}, with gray lines being LSWT energies for the parameter set~\#12 (\ref{eq_SM_set12}). Solid colored lines are from set (\ref{eq_set_Jpp}), Fig.~\ref{fig_INS}. }
\label{FigSM:set12_fit}
\vskip -0.5cm
\end{figure}

The LSWT energies using set~\#12~(\ref{eq_SM_set12}) provide an exceptionally good fit to {\it all} available INS data for all  fields   in the polarized phase of BaCAO, with Fig.~\ref{FigSM:set12_fit} showing  most of these results (gray lines) together with the fits using the set (\ref{eq_set_Jpp}) from Fig.~\ref{fig_INS}. These two sets provide largely indistinguishable results for the data in Figs.~\ref{fig_spectrum}(a), (b) and \ref{FigEM:H_075T_INS}(a), (b).

However, while set~\#12 provides  better agreement with the INS data than set (\ref{eq_set_Jpp}), especially at lower fields, DMRG calculations suggest that it is in a ZZ ground state at zero field, placing it outside the dZZ phase in  Fig.~\ref{fig_spectrum}(c). This indicates that some fluctuations may still be affecting the spectra at lower fields, and that the ideal LSWT fit in this regime may be somewhat misleading. 

By exploring the phase diagram of the model with the help of the DMRG scans discussed in the EM of the main text and with the set $\{\}^\diamond$ of Eq.~(\ref{eq_set8_Jpp}), marked by an empty diamond in Fig.~\ref{fig_spectrum}(c), we have arrived at the set (\ref{eq_set_Jpp}) as our ultimate choice that resolves most of the enigmas of BaCAO. Regardless of this final choice, the large values of ${\sf J}_{z\pm}$, and correspondingly dominant values of the Kitaev exchange (\ref{eq_set}),  are an inescapable conclusion of our study. 

Upon finalizing our strategy, we realized that the prior $XXZ$ modeling of BaCAO in Refs.~\cite{LP84,LP90} has  significant similarities to the $XXZ$ sector of our ultimate set (\ref{eq_set_Jpp}), further validating our resultant choice by providing a  simpler, more intuitive way of arriving at the same point in the multi-parameter space.

% ==============================================================================
% ==============================================================================

\vspace{-0.3cm}
% ==============================================================================
\section{DMRG field scans and non-scans}
% ==============================================================================
\vskip -0.2cm

%\vspace{-0.3cm}
% ==============================================================================
\subsection{Scans}
% ==============================================================================
\vskip -0.2cm

The DMRG scans vs various parameters  were used to determine the extent of the dZZ and other intermediate phases between the FM and ZZ, as described in the main text and EM; see the corresponding discussions of Figs.~\ref{fig_spectrum}(c) and \ref{FigEM:Scans}. In addition to these scans, and  to provide further support for the proposed parameter set for BaCAO, we have also performed DMRG scans with both in-plane and out-of-plane fields. We also note that a similar analysis has   recently been performed in Ref.~\cite{aRu_saga} for the $\alpha$-RuCl$_3$ model by some of the same authors.

We performed DMRG scans vs field in the two principal in-plane directions, $a^*$ and $b$ ($x$ and $y$), parallel and perpendicular to the bond, respectively, and for the out-of-plane field along the $z$ axis, all using the parameter set~\eqref{eq_set_Jpp}. We also used the same in-plane g-factors in DMRG scans, $g_{a*}\!=\!g_b\!=\!4.8$ for simplicity, and the out-of-plane g-factor $g_c\!=\!2.4$~\cite{LP90,Broholm_BCAO}; $h_\alpha\!=\!g_\alpha \mu_B B_\alpha$.

\begin{figure}[t]
\includegraphics[width=\linewidth]{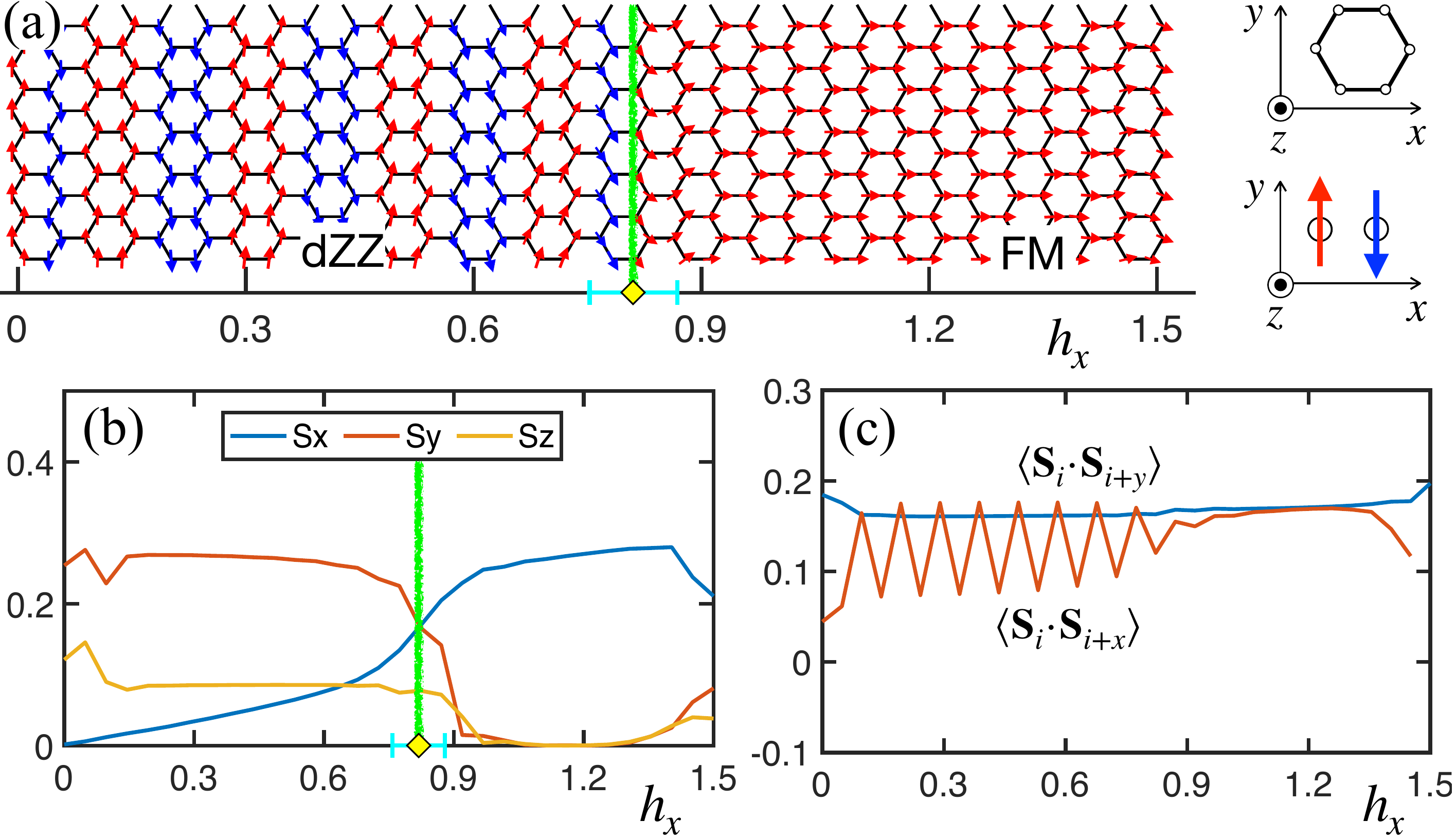}
\vskip -0.2cm
\caption{(a) The $32\!\times\!12$ DMRG scan for the parameter set~\eqref{eq_set_Jpp}  vs field along $x$. The arrows are the ordered moments' projections onto the $xy$ plane. (b) and (c)  show three components of the on-site ordered moment, $\langle S_i^\alpha \rangle$,   and  the nearest-neighbor correlators $\langle {\bf S}_i \!\cdot\! {\bf S}_{i+x(y)} \rangle$, respectively, averaged over the vertical columns. Transitions to the polarized phase are indicated.}
\label{FigSM:Scan_x}
\vskip -0.5cm
\end{figure}

The results for  $\mathbf{B} \!\parallel \!x\! \parallel\! a^*$ are presented in Fig.~\ref{FigSM:Scan_x}. In Fig.~\ref{FigSM:Scan_x}(a), the presentation of spins is similar to that in Fig.~\ref{FigEM:Scans}, with the key difference being that the spin-projection plane is now $xy$ in order to emphasize the polarizing effect of the field.  

% ==============================================================================
% ==============================================================================

Fig.~\ref{FigSM:Scan_x}(b) shows the evolution of the three components of the on-site ordered moment, $\langle S_i^\alpha \rangle$, averaged over the vertical direction (circumference of the cylinder), vs field. The complementary Fig.~\ref{FigSM:Scan_x}(c) shows the same for the nearest-neighbor correlators $\langle {\bf S}_i \!\cdot\! {\bf S}_{i+x} \rangle$ and $\langle {\bf S}_i \!\cdot\! {\bf S}_{i+y} \rangle$ averaged the same way. The transition to the polarized phase and its error bars are determined from the inflection points in the ordered moment curves and the width of the transition region, respectively.

One can see a gradual transition from the dZZ state in zero field to a polarized state, with the inflection point in the magnetization plot in Fig.~\ref{FigSM:Scan_x}(b) suggesting a critical field in the range of  $0.75$--$0.85$~T.

% ==============================================================================
\begin{figure}[t]
\includegraphics[width=\linewidth]{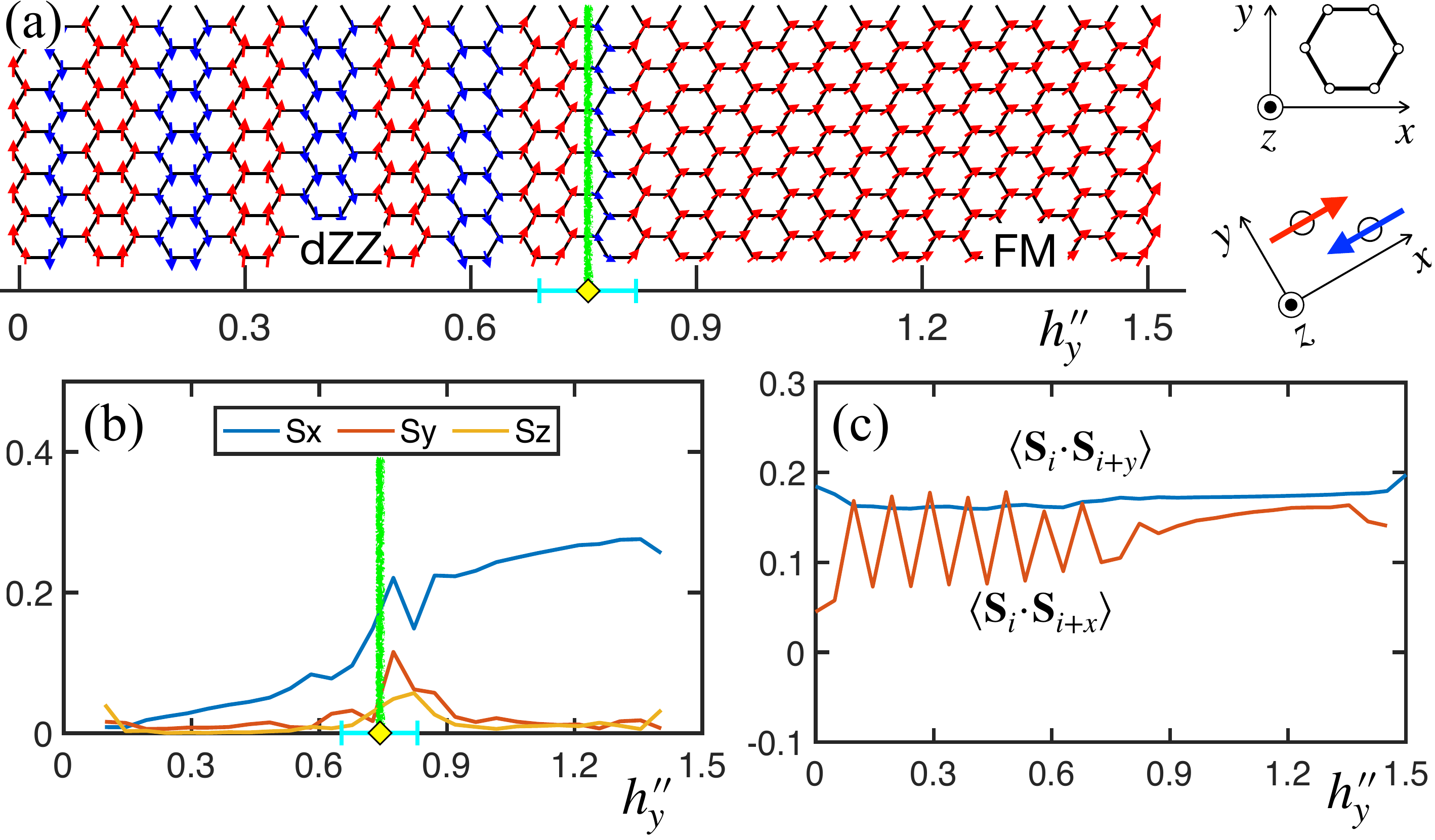}
\vskip -0.2cm
\caption{Same as in Fig.~\ref{FigSM:Scan_x} for $\mathbf{B}\!\parallel\! y$; axes for spins are rotated to align $x'$ axis with the field, see text.}
\vskip -0.4cm
\label{FigSM:Scan_y}
\end{figure}
% ==============================================================================

Similarly, Fig.~\ref{FigSM:Scan_y} shows a scan for the field $\mathbf{B} \!\parallel\! b$. Due to the selection of a specific dZZ domain by DMRG, we must use a tilted direction of the field, at a $\pi/6$ angle relative to the cylinder $x$-axis, to have it perpendicular to the bond, with the spin axes tilted accordingly. Because of the tilt, the averaging of the spin components $\langle S_i^\alpha \rangle$ and correlators $\langle {\bf S}_i \!\cdot\! {\bf S}_{i+x} \rangle$ and $\langle {\bf S}_i \!\cdot\! {\bf S}_{i+y} \rangle$ was done over the two next-nearest vertical zigzag columns to minimize the oscillatory trends in these quantities.

As one can see in Fig.~\ref{FigSM:Scan_y}(b), an inflection point around $0.7(1)$~T suggests a transition to the polarized state.

These results are in close agreement with the low critical in-plane fields observed for BaCAO, supporting the overall coherence of the phenomenological outcomes from the  parameter set proposed in this work.

% ==============================================================================
\begin{figure}[b]
%\vskip -0.3cm
\includegraphics[width=\linewidth]{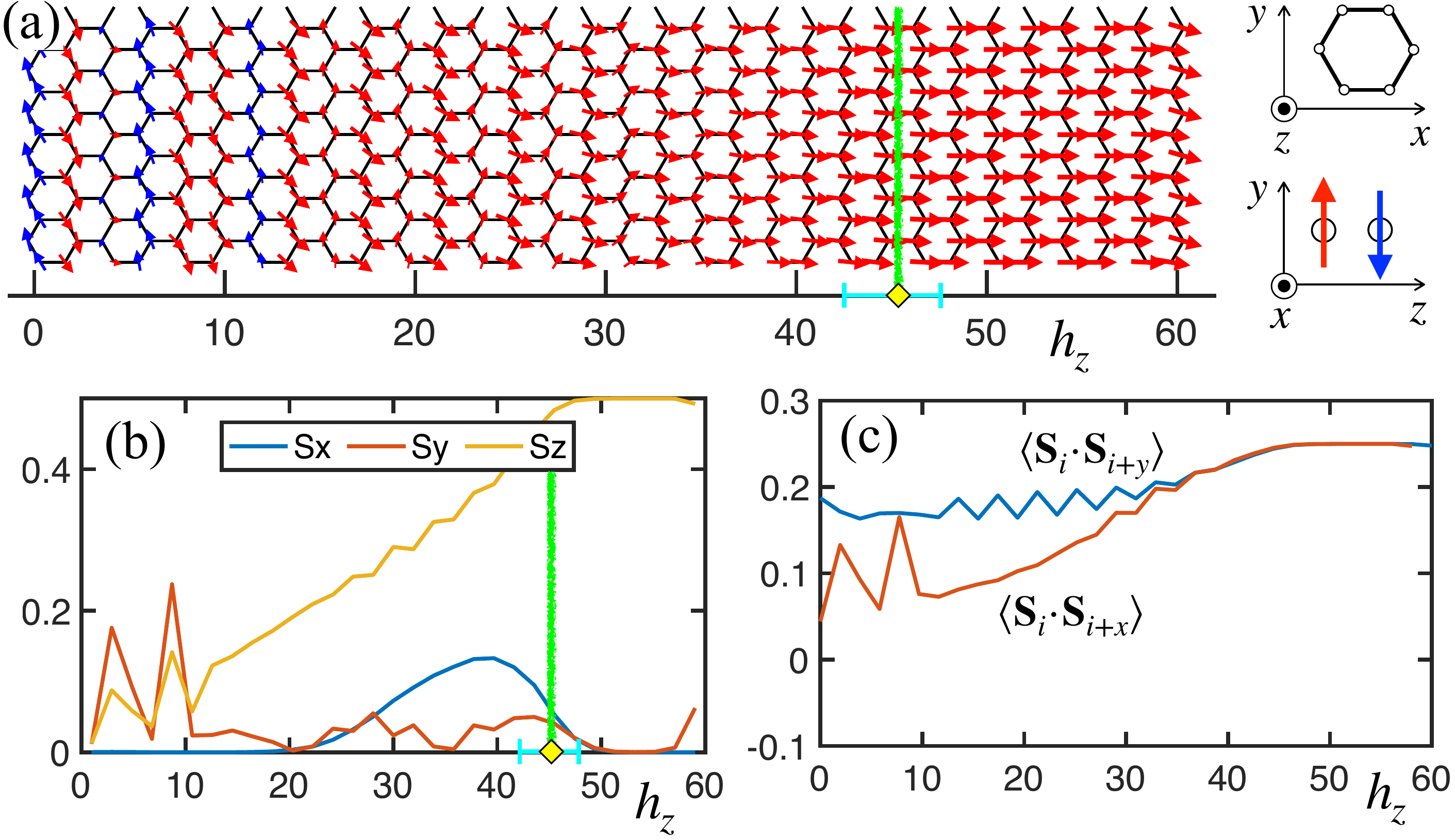}
\vskip -0.2cm
\caption{Same as in Fig.~\ref{FigSM:Scan_x} for $\mathbf{B}\!\parallel\! z$. In (a), spins are shown in the $yz$ plane.}
\label{FigSM:Scan_z}
\end{figure}
% ==============================================================================

Last, but not least, we also performed a DMRG  scan with the field normal to the honeycomb plane. In Fig.~\ref{FigSM:Scan_z}(a), magnetic moments are shown in the $yz$ plane. Together with the ordered moment components in Fig.~\ref{FigSM:Scan_z}(b) and the nearest-neighbor correlators in Fig.~\ref{FigSM:Scan_z}(c), one can see that this scan suggests a more complicated magnetization process with a potential intermediate phase or phases. The most stark difference from the in-plane field results is the actual magnitude at which  full saturation along $z$ is achieved. With $g_c/g_{ab}\!\approx\!0.5$~\cite{LP90,Broholm_BCAO} and significant $XXZ$ anisotropy $\Delta\!\approx\!0.4$, the two orders of magnitude difference is still impressive. The value of $B_c^{(z)}\!\approx\!47(3)$~T from DMRG is also closely compatible with  recent experiments~\cite{Zapf_2024}, further supporting the validity of our model.

\vspace{-0.3cm}
% ==============================================================================
\subsection{UUD phase, non-scan}
% ==============================================================================
\vskip -0.2cm

% ==============================================================================
\begin{figure}[t]
\includegraphics[width=\linewidth]{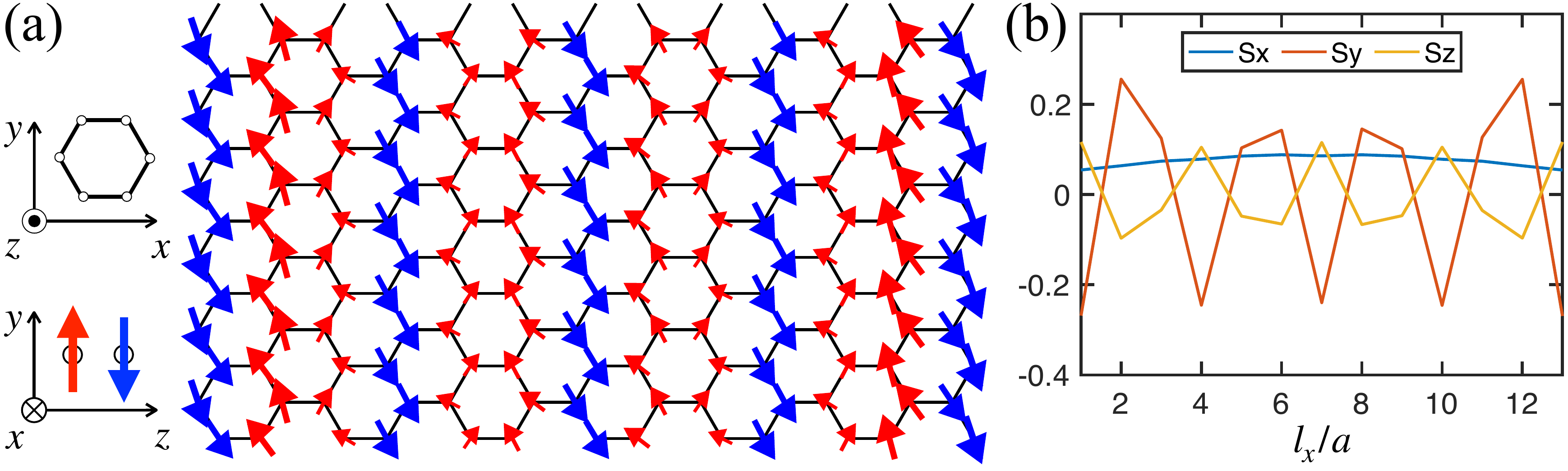}
\vskip -0.2cm
\caption{(a) The UUD  state of the model (\ref{eq_set_Jpp}) in a $12\times 13$ XC cluster in a magnetic field $h_x\!=\!0.5$~T along the $x$ direction; spins are shown in the $yz$ plane. (b) The spin components averaged over the vertical columns.}
\label{FigSM:UUD}
\vskip -0.5cm
\end{figure}
% ==============================================================================

One outstanding feature of BaCAO's in-plane-field phase diagram is the intermediate, field-induced ``up-up-down'' (UUD) columnar state, which can be viewed as a mixture of the dZZ and ZZ phases~\cite{Tsirlin_BCAO_2024,LP90}. This UUD state extends from  $\sim\!0.2$~T up to the saturation field.  

The small energy scale of these fields---and the experimental observation that a modest pressure suppresses the UUD phase~\cite{Budko_2022,Tsirlin_BCAO_2024}---point to a near-degeneracy among different arrangements of ferromagnetic zigzag chains, coupled either parallel or antiparallel.  Such near-degeneracy is reminiscent of the {\it classical} $XXZ$ FM-AFM  $J_1$--$J_3$ model at $J_3/|J_1|\!=\!1/3$, where FM, ZZ, dZZ, and other multiple-zigzag  states coincide in energy, and of prior quantum studies that found higher-zigzag states (e.g. triple-zigzag) in the $J_1$--$J_2$--$J_3$ $XXZ$ model~\cite{shengtao_j1j3}, as well as of our quantum phase diagram in Fig.~\ref{fig_spectrum}(c).

At first glance, the field scans in Figs.~\ref{FigSM:Scan_x} and \ref{FigSM:Scan_y} do not show an intermediate UUD phase.  However, this absence is likely because of the gradient of field along the scan and the large unit cell of the UUD pattern.  

To confirm the UUD state directly, we performed a non-scan DMRG calculation on a $12\times 13$ XC cluster, which can accommodate several UUD unit cells lengthwise, in a representative 0.5~T field applied along $x$.  Our Fig.~\ref{FigSM:UUD} provides a direct evidence of the UUD state realized for our BaCAO model (\ref{eq_set_Jpp}). Fig.~\ref{FigSM:UUD}(a) shows  the $yz$ plane of the spin projections, the nearly uniform $S^x$ component in Fig. \ref{FigSM:UUD}(b) is field-induced, while the $y$ and $z$ components form a clear three-column UUD structure that also has a finite out-of-plane tilt.  

We also find that the UUD state in this cluster is nearly degenerate with the ZZ and FM states, underscoring its delicate nature.  Altogether, these results provide direct evidence that our BaCAO model captures the most, if not all,  key phenomenological features of the material, including its intermediate UUD columnar phase.

\vspace{-0.3cm}
% ==============================================================================
\section{Further details of the analysis}
% ==============================================================================
\vskip -0.2cm

One may notice that for the anisotropic-exchange terms in our proposed parameter set in Eq.~(\ref{eq_set_Jpp}) and in all other parameter sets used in our additional analyses, such as in Eq.~(\ref{eq_set8_Jpp}) and Eq.~(\ref{eq_SM_set12}),  the $\sf{J}_{z\pm}$ term is significant while the $\sf{J}_{\pm\pm}$ term is small. The $\sf{J}_{\pm\pm}$ term is also small in the previous analysis of Ref.~\cite{Broholm_BCAO} where $\sf{J}_{z\pm}$ was neglected.

Both terms contribute linearly to the value of the Kitaev exchange, $K\!=\!\sqrt{2}{\sf J}_{z\pm}\!-2{\sf J}_{\pm\pm}$; see the note below Eq.~(\ref{eq_HJpm}) in the main text and the transformation matrix in the EM, Eq.~(\ref{KJGG1_model}). However, their contributions to the magnon energy splitting at the M point for the two  principal in-plane field directions in the polarized phase, Eq.~(\ref{eq_EM_Ediff}), are different. The term $\sf{J}_{z\pm}$ enters it as a square, while  $\sf{J}_{\pm\pm}$ enters linearly, multiplied by a combination of (large) $XXZ$ exchanges. Thus, small changes in $\sf{J}_{\pm\pm}$ need to be compensated by larger changes in $\sf{J}_{z\pm}$ to maintain the same experimentally observed energy splitting at the M point. 

This raises a possibility that a different combination of $\sf{J}_{\pm\pm}$ and $\sf{J}_{z\pm}$ terms can produce the same, or similar spectra in the polarized phase, and correspond to smaller values of the Kitaev term by trading a large $\sf{J}_{z\pm}$ term for a smaller $\sf{J}_{\pm\pm}$ term.

% ==============================================================================
\begin{figure}[t]
\includegraphics[width=\linewidth]{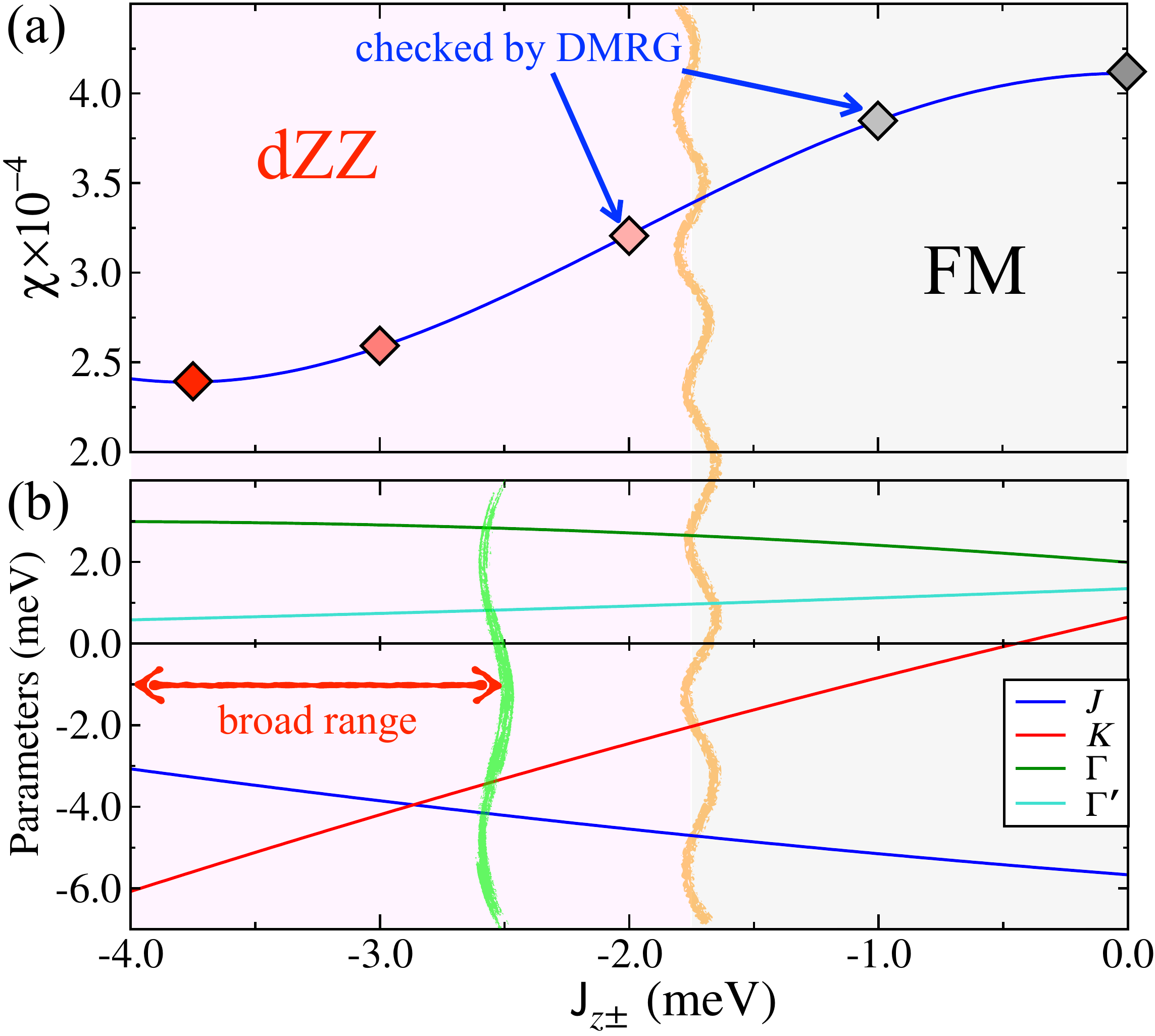}
\vskip -0.2cm
\caption{(a)  Quality fit factor (\ref{eq_SM_chi}) vs $\sf{J}_{z\pm}$. Symbols are sets checked by DMRG, with the phase boundary inferred.  (b) $\{J,K,\Gamma,\Gamma'\}$  sets vs $\sf{J}_{z\pm}$, possible broad range; see text.}
\label{FigSM:chi}
\vskip -0.5cm
\end{figure}
% ==============================================================================

We have investigated this possibility by combining the analysis of the quality fit factor, or the fit error function, and DMRG calculations for several additional values of parameters.
The brief summary of this investigation is presented in Fig.~\ref{FigSM:chi}. The first panel, Fig.~\ref{FigSM:chi}(a), shows the quality fit factor as a function of $\sf{J}_{z\pm}$. It is defined as
\vskip -0.15cm
\noindent
\begin{align}
\label{eq_SM_chi}
\chi=\sum_i \frac{\big( E_i-E_{\mathbf{q}_i}\big)^2}{\delta E_i^2},
\end{align}
\vskip -0.15cm
\noindent
where $E_i$ and $\delta E_i$ are the positions and errors of the magnon peaks obtained from raw neutron scattering data using Gaussian fits, and $E_{\mathbf{q}_i}$ are the values of the magnon energies from LSWT, see Sec.~\ref{Sec_LSWT}, calculated for experimental transfer momentum values $\mathbf{q}_i$.

Note that for each value of $\sf{J}_{z\pm}$ along the horizontal axis, all parameters of the model are different and are extracted according to the procedure described in Sec.~\ref{Sec_INS_LSWT_parameters} with $\sf{J}_{z\pm}$ fixed. The quality fit factor is obtained using all available INS data in the polarized phase, shown in Figs.~\ref{fig_lattice}, \ref{fig_INS}, \ref{fig_spectrum}, and \ref{FigEM:H_075T_INS}.

We also underscore that the parameter set used in this work to describe BaCAO, Eq.~\eqref{eq_set_Jpp}, was {\it not} obtained using the quality fit factor given in (\ref{eq_SM_chi}), but instead used the approach described in the main text and Sec.~\ref{Sec_INS_LSWT_parameters}. However, the provided agreement between theoretical magnon energies and data in Figs.~\ref{fig_INS}, \ref{fig_spectrum}, and \ref{FigEM:H_075T_INS} suggests that the fitting error for this parameter set must be small.  

One can see in Fig. \ref{FigSM:chi}(a) that the value of ${\sf J}_{z\pm}\!=\!-3.76$~meV from the set \eqref{eq_set_Jpp} is, indeed, very near the minimum of the quality fit factor curve. As we move away from that minimum, we sacrifice the quality of the INS data fits by the theory quite significantly, but it mainly affects such fits at the lower fields, approaching the critical fields from above.

The lower panel, Fig.~\ref{FigSM:chi}(b), shows the corresponding parameter changes of the nearest-neighbor exchange matrix in the generalized Kitaev-Heisenberg parameterization. One can see that the terms that are affected the strongest by the change of $\sf{J}_{z\pm}$ are the $J$ and the $K$ terms, while $\Gamma$ and $\Gamma'$ are affected significantly less. In agreement with our expectations, the absolute value of the $K$ term is decreased nearly linearly with the decrease of $\sf{J}_{z\pm}$ term, while the absolute value of the $J$ term increases.

The DMRG input is essential here. It shows that the phase boundary between the dZZ and the FM phases along the ${\sf J}_{z\pm}$ axis is between $-2$~meV and $-1.5$~meV. 

The averaged out-of-plane tilt angle for the magnetic moments for ${\sf J}_{z\pm}\!=\!-3$~meV is $7.3\degree$ and the in-plane critical fields are about 0.7~T. At ${\sf J}_{z\pm}\!=\!-1$~meV, the ground state is firmly FM. At ${\sf J}_{z\pm}\!=\!-2$~meV, the ground state is still the same type of the dZZ state that is described in the main text with the corresponding out-of-plane tilt angle about $6.3\degree$, which is also compatible with the neutron polarization data. However, the in-plane critical fields are $\sim \!0.35$~T, significantly lower than the experimental ones. Given the finite-size effects in DMRG, it suggests that this value of $\sf{J}_{z\pm}$ is already outside of the physically allowed range. 

Therefore, we mark the region between $-4$~meV and $-2.5$~meV for $\sf{J}_{z\pm}$ as a broadly plausible range for the bond-dependent terms to be adjusted to, at the price of much greater fitting error, shown in Fig.~\ref{FigSM:chi}(a). 

Although there is presently no reason for us to assume such a range, one can allow it as a wider possibility. However, even at the smaller $|{\sf J}_{z\pm}|$, the corresponding absolute values of the Kitaev term are still large, and, even if it is not the dominant term anymore, it is still on par with the exchanges $J$ and $\Gamma$, thus completely excluding the possibility of a small Kitaev term model as a relevant description of the physics of BaCAO.

\vspace{-0.3cm}
% ==============================================================================
\section{Other terms and parameter sets}
% ==============================================================================
\vskip -0.2cm

We briefly comment on two additional issues. 

It was noted that the lower symmetry, which results in a small ``puckering'' of Co$^{2+}$ ions in and out of the honeycomb plane, allows for the presence of additional terms in the exchange matrix in BaCAO~\cite{Broholm_BCAO}. These terms may be responsible for a small  in-plane tilt of $2.4(1.0)\degree$ of the magnetic moments within the double-zigzag structure observed experimentally in the neutron polarization analysis of Ref.~\cite{LP18}, but not captured by our model.

Recently, a manuscript dedicated to the study of BaCAO's magnetic state, excitations, and model has been
submitted; see Ref.~\cite{Songvilay25}. Their experimental data and analysis support the presence of strong Kitaev-like terms in BaCAO's model and provide important insights into potentially more complex physics of
BaCAO's intermediate phases.

{\it Note: All references in this Supplemental Material refer to the references listed in the main text.}

\end{document}